\documentclass[10pt,aps,prc,floatfix,twocolumn,nofootinbib]{revtex4-1}

\usepackage[labelfont={small},subrefformat=parens,caption=false]{subfig}
\usepackage{graphicx,amsmath,amssymb,bm}
\usepackage{amsfonts}
\usepackage[utf8]{inputenc}
\usepackage{verbatim}
\usepackage{float}
\usepackage{cancel}
\usepackage{multirow}
\usepackage{array}
\usepackage{xparse}

\usepackage{physics}
\usepackage{color}
\usepackage{dsfont}

\usepackage[pdfpagelabels]{hyperref}

\usepackage[overload]{textcase}

\usepackage[linesnumbered,algoruled,lined,commentsnumbered]{algorithm2e}

\usepackage{cellspace}
\graphicspath{{./figures/}}

\usepackage{silence}
\usepackage[T1]{fontenc}

\makeatletter
\newcommand\newsubcommand[3]{\newcommand#1{#2\sc@sub{#3}}}
\def\sc@sub#1{\def\sc@thesub{#1}\@ifnextchar_{\sc@mergesubs}{_{\sc@thesub}}}
\def\sc@mergesubs_#1{_{\sc@thesub#1}}

\newcommand\newsupcommand[3]{\newcommand#1{#2\sc@sup{#3}}}
\def\sc@sup#1{\def\sc@thesup{#1}\@ifnextchar^{\sc@mergesups}{^{\sc@thesup}}}
\def\sc@mergesups^#1{^{\sc@thesup#1}}
\makeatother

\DeclareMathAlphabet{\mathbcal}{OMS}{cmsy}{b}{n}

\newcommand{\eqrevisited}[1]{\tag{\ref*{#1}$^\prime$}}

\DeclareMathOperator{\ninvchisq}{N\chi^{--2}}  % Normal Inverse chi2
\DeclareMathOperator{\invchisq}{\chi^{--2}}  % Inverse chi2
\DeclareMathOperator{\E}{\mathbb{E}}

\newcommand{\kernel}{\kappa}
\newcommand{\onevec}{\boldvec{1}}
\newcommand{\identity}{\mathds{I}}
\newcommand{\sdth}{\bar c}

\newcommand{\sdnugget}{\sigma_n}

\newcommand{\diagnostic}{\textup{D}}
\DeclareMathOperator{\CI}{CI}
\newcommand{\DCI}{\diagnostic_{\CI}}
\newcommand{\MD}{\textup{MD}}
\newcommand{\DMD}{\diagnostic_{\MD}}
\newcommand{\DVAR}[1]{\inputvec{D}_{\textup{#1}}}
\newcommand{\DVARIT}[1]{\inputvec{D}_{#1}}

\newcommand{\iid}{\text{i.i.d.}}

\newcommand{\indicator}{\mathbf{1}}

\newcommand{\hypm}{\eta}
\newcommand{\hypdisp}{V}
\newcommand{\hypdf}{\nu}
\newcommand{\hyptau}{\tau}

\newcommand{\basisfunc}{b}
\newcommand{\basis}{B}

\newcommand{\ftrain}{\inputvec{f}_{\textup{train}}}
\newcommand{\fvalid}{\inputvec{f}_{\textup{val}}}

\newcommand{\genobsmean}[1]{m_{#1}}
\newcommand{\genobscorr}[1]{R_{#1}}

\newcommand{\genobsbasisfunctrans}[1]{\basisfunc_{#1}}  % ^\trans
\newcommand{\tildegenobsbasisfunc}[1]{\tilde\basisfunc_{#1}}

\newcommand{\tildegenobsmean}[1]{\tilde m_{#1}}
\newcommand{\tildegenobscorr}[1]{\tilde R_{#1}}

\newcommand{\discrmean}[1]{m_{\delta #1}}
\newcommand{\discrcorr}[1]{R_{\delta #1}}
\newcommand{\discrbasisfunc}[1]{\basisfunc_{\delta #1}}
\newcommand{\discrbasisfunctrans}[1]{\basisfunc_{\delta #1}}  % ^\trans

\newcommand{\tildediscrbasisfunc}[1]{\tilde\basisfunc_{\delta #1}}

\newcommand{\tildediscrmean}[1]{\tilde m_{\delta #1}}
\newcommand{\tildediscrcorr}[1]{\tilde R_{\delta #1}}

\newcommand{\hypprior}[1]{#1_0}
\newcommand{\cond}[1]{#1}

\newcommand{\NkLO}[1]{\ensuremath{\mathrm{N}^{#1}\mathrm{LO}}}

\newcommand{\nc}{n_c}

\newcommand{\data}{\mathcal{D}}

\newcommand{\param}{\boldsymbol{\theta}}

\DeclareMathOperator{\GP}{\mathcal{GP}}
\DeclareMathOperator{\TP}{\mathcal{TP}}

\newcommand{\npr}{\ensuremath{np}}

\newcommand{\boldvec}[1]{\bm{#1}}

\newcommand{\ordervec}{\vec}
\newcommand{\inputdimvec}{}
\newcommand{\inputvec}{\mathbf}
\newcommand{\func}{f}
\newcommand{\funcvec}{\inputvec{\func}}
\newcommand{\meanvec}{\inputvec{m}}

\newcommand{\muth}{m_{\textup{th}}}
\newcommand{\muthset}{\mathbf{m}_{\textup{th}}}
\newcommand{\covth}{\Sigma_{\textup{th}}}
\newcommand{\covexp}{\Sigma_{\textup{exp}}}
\newsubcommand{\ckvec}{\ordervec{c}}{k}
\newcommand{\ckvecsq}{\ckvec^{\,2}}

\newsubcommand{\bkvec}{\ordervec{b}}{k}
\newcommand{\kinparvec}{\inputdimvec{x}}
\newcommand{\kinparvecset}{\inputdimvec{\inputvec{x}}}

\newsubcommand{\ckvecset}{\ordervec{\inputvec{c}}}{k}

\newsubcommand{\ckvecapprox}{\mathbf{c}'}{k}
\newsubcommand{\ckvecapproxset}{\mathbf{C}'}{k}

\newsubcommand{\bkvecapprox}{\mathbf{b}'}{k}
\newsubcommand{\bkvecset}{\mathbf{B}}{k}
\newsubcommand{\bkvecapproxset}{\mathbf{B}'}{k}

\newcommand{\genobs}{y}

\newsubcommand{\genobsvec}{\ordervec{\genobs}}{k}
\newsubcommand{\genobsvecset}{\ordervec{\inputvec{\genobs}}}{k}
\newcommand{\genobsset}{\inputvec{\genobs}}

\newcommand{\genobsexp}{\genobs_{\textup{exp}}}        % subscript or superscript?
\newcommand{\genobsexpset}{\inputvec{\genobs}_{\textup{exp}}}
\newcommand{\genobsth}{\genobs_{\textup{th}}}          % subscript or superscript?

\newcommand{\lecs}{\vec{a}}

\newsubcommand{\akvec}{\mathbf{a}}{k}

\newsubcommand{\akvecapprox}{\mathbf{a}'}{k}
\newsubcommand{\akvecset}{\mathbf{A}}{k}
\newsubcommand{\akvecapproxset}{\mathbf{A}'}{k}

{}  % Remove the definition from the Physics package

\DeclareMathOperator{\pr}{pr} % Good, but want to handle sizing and | spacing?
\newcommand{\given}{\,|\,}  % Use for | in \pr

\newcommand{\Elab}{E_{\rm lab}}

\newcommand{\prel}{p_{\rm rel}}

\newcommand{\degr}{\circ}

\newcommand{\normal}{\mathcal{N}}

\newcommand{\transpose}[1]{{#1}^{\intercal}}
\newcommand{\trans}{\intercal}

\newcommand{\genobsref}{\ensuremath{y_{\mathrm{ref}}}}

\newcommand{\sigmatot}{\sigma_{\text{tot}}}

\def\diffd{\mathrm{d}}  % Upright differentials
\DeclareDocumentCommand\differential{ o g d() }{ % Differential 'd'
    \IfNoValueTF{#2}{
        \IfNoValueTF{#3}
            {\diffd\IfNoValueTF{#1}{}{^{#1}}}
            {\mathinner{\diffd\IfNoValueTF{#1}{}{^{#1}}\argopen(#3\argclose)}}
        }
        {\mathinner{\diffd\IfNoValueTF{#1}{}{^{#1}}#2} \IfNoValueTF{#3}{}{(#3)}}
    }
\DeclareDocumentCommand\dd{}{\differential} % Shorthand for \differential

\newcommand{\pathd}{\mathcal{D}}  % differential symbol for path integrals

\DeclareDocumentCommand\pathdifferential{ o g d() }{ % Path 'D'
    \IfNoValueTF{#2}{
        \IfNoValueTF{#3}
            {\pathd\IfNoValueTF{#1}{}{^{#1}}}
            {\mathinner{\pathd\IfNoValueTF{#1}{}{^{#1}}\argopen(#3\argclose)}}
        }
        {\mathinner{\pathd\IfNoValueTF{#1}{}{^{#1}}#2} \IfNoValueTF{#3}{}{(#3)}}
    }

\begin{document}

% Overall checks and general comments

\title{Quantifying Correlated Truncation Errors in Effective Field Theory}

\author{J.~A. Melendez}
\email{melendez.27@osu.edu}
\affiliation{Department of Physics, The Ohio State University, Columbus, OH 43210, USA}

\author{R.~J. Furnstahl}
\email{furnstahl.1@osu.edu}
\affiliation{Department of Physics, The Ohio State University, Columbus, OH 43210, USA}

\author{D.~R.~Phillips}
\email{phillid1@ohio.edu}
\affiliation{Department of Physics and Astronomy and Institute of Nuclear and Particle Physics, Ohio University, Athens, OH 45701, USA}
\affiliation{Institut f\"ur Kernphysik, Technische Universit\"at Darmstadt, 64289 Darmstadt, Germany}
\affiliation{ExtreMe Matter Institute EMMI, GSI Helmholtzzentrum f{\"u}r Schwerionenforschung GmbH, 64291 Darmstadt, Germany}

\author{M.~T. Pratola}
\email{mpratola@stat.osu.edu}
\affiliation{Department of Statistics, The Ohio State University, Columbus, OH 43210, USA}

\author{S.~Wesolowski}
\email{scwesolowski@salisbury.edu}
\affiliation{Department of Mathematics and Computer Science, Salisbury University, Salisbury, MD 21801, USA}

\date{\today}

\begin{abstract}
Effective field theories (EFTs) organize the description of complex systems into an infinite sequence of decreasing importance.
Predictions are made with a finite number of terms, which induces a truncation error that is often left unquantified.
We formalize the notion of EFT convergence and propose a Bayesian truncation error model for predictions that are correlated across the independent variables, e.g., energy or scattering angle.
Central to our approach are Gaussian processes that encode both the naturalness and correlation structure of EFT coefficients.
Our use of Gaussian processes permits efficient and accurate assessment of credible intervals, allows EFT fits to easily include correlated theory errors, and provides analytic posteriors for physical EFT-related quantities such as the expansion parameter.
We demonstrate that model-checking diagnostics---applied to the case of multiple curves---are powerful tools for EFT validation. As an example, we assess a set of nucleon-nucleon scattering observables in chiral EFT\@.
In an effort to be self contained, appendices include thorough derivations of our statistical results.
Our methods are packaged in Python code, called \texttt{gsum}~\cite{gsum}, that is available for download on GitHub.
\end{abstract}

\maketitle

\newpage

%%%%%%%%%%%%%%%%%%%%%%%%%%%%%%%%%%%%%%%%%%%%%%%%%%%%%%%%%%%%%%%

\section{Introduction}
\label{sec:introduction}

The \emph{power counting} of an effective field theory (EFT) mandates how to organize an infinite number of operators into a sequence of decreasing importance.
The precision of EFT predictions is then governed by the uncertainty in the following quantities: (1) the fit of the parameters, or low energy constants (LECs), of the EFT, (2) the error due to truncation of the infinite EFT series, and (3) other approximations made in prediction.
For chiral EFT calculations, recent advances in precision motivate a rigorous accounting of all these theoretical uncertainties, but such an accounting is desirable for EFTs across all domains.
This work will focus on the uncertainty due to truncation, but our results have consequences for
how LECs should be fit~\cite{WesolowskiExploringBayesianparameter2019}.

In previous works, we proposed a pointwise Bayesian statistical model to estimate EFT truncation errors for predicted observables $\genobs$~\cite{furnstahl_quantifying_2015,MelendezBayesiantruncationerrors2017} (see also Ref.~\cite{cacciari_meaningful_2011}).
This model formalizes the notion of convergence in $\genobs$, which allows one to credibly assert which observations are consistent with theory.
One can incorporate expert knowledge on the convergence pattern (inherited from the EFT power counting) into prior distributions, and subsequently update these beliefs given order-by-order predictions $\{\genobs_n\}$~\cite{cacciari_meaningful_2011}.
If the observable is itself a function $\genobs(\kinparvec)$, e.g., a cross section at a range of energies, Refs.~\cite{furnstahl_quantifying_2015,MelendezBayesiantruncationerrors2017} simply compute its posterior in a pointwise manner.
This pointwise model is tractable, but it is flawed since it ignores information about $\genobs(\kinparvec)$ as a whole, that is, the correlations of $\genobs$ at nearby $\kinparvec$.

Here we extend the pointwise model to functions $\genobs(\kinparvec)$,
encoding the idea of curvewise convergence for observables via Gaussian processes (GPs).
GPs are powerful tools for both regression and classification, and have become popular in statistics, physics, applied mathematics, machine learning, and geostatistics~\cite{sacks1989design,cressie1992statistics,rasmussen_gaussian_2006}.
Their popularity is due in part to their modeling flexibility and the mathematical convenience of Gaussian distributions.
Despite the leap from points to curves, our algorithm remains analytic due to a convenient yet flexible choice of priors.
The GP parameters are interpretable from an EFT convergence standpoint, and can be easily \emph{calibrated} against known order-by-order predictions.

We develop the intuition behind our convergence model and provide a primer on GPs (Sec.~\ref{sec:the_model}), and then explore ways to assess whether our assumptions are failing (Sec.~\ref{sec:model_checking}).
To display the procedure in action, we show success and failure using toy EFT predictions (Sec.~\ref{sec:toy_application}) and assess nucleon-nucleon ($NN$) scattering predictions in chiral EFT (Sec.~\ref{sec:NN_scattering}).
We then discuss future prospects and conclude the main text (Sec.~\ref{sec:summary}) before providing an extended discussion on the statistical derivations relevant to the pointwise and GP models (Appendix~\ref{app:derivations}).

%%%%%%%%%%%%%%%%%%%%%%%%%%%%%%%%%%%%%%%%%%%%%%%%%%%%%%%%%%%%%%%

\section{The Model}
\label{sec:the_model}

\subsection{Setup}
\label{sub:setup}

Predictions of experimental quantities can be inaccurate due to shortcomings of the theoretical framework and noisy measurements.
This work focuses on building a model discrepancy term for EFT truncation errors.
The exact manner by which specific errors propagate can be complicated, but here we assume a simple additive model of theoretical discrepancy,
\begin{align} \label{eq:discrepancy_model}
    \genobsexp(\kinparvec) = \genobsth(\kinparvec;\lecs) + \delta\genobsth(\kinparvec) + \delta\genobsexp(\kinparvec),
\end{align}
where $\kinparvec \in \mathbb{R}^d$ is the independent variable,%
%\footnote{For example, a dimension $d=2$ could correspond to a $NN$ spin observable as a function of energy and scattering angle.}, 
% \footnote{A dimension $d=2$ example would be a $NN$ observable that is a function of energy and scattering angle.},
\footnote{The differential cross section, as a function of both energy and scattering angle, is an example where $d=2$.}
$\lecs$ are the fitted parameters of the theory prediction $\genobsth$ (here, the LECs of an EFT\footnote{Note that the LECs $\lecs$ are \emph{not} the same as the observable coefficients $c_n$ introduced in Eq.~\eqref{eq:Xk}.
These parameters $\lecs$ do not play an explicit role in our treatment of truncation errors, although they do appear via a brief discussion of EFT parameter fitting in Sec.~\ref{sub:application_types}.}), and $\genobsexp$ are experimental measurements.
The uncertainties associated with the theory and experiment are given by $\delta\genobsth$ and $\delta\genobsexp$.
Of course, if the exact values of $\delta\genobsth$ and $\delta\genobsexp$ were known, then there would be no uncertainty.
Instead, $\delta\genobsth$ and $\delta\genobsexp$ are treated as random variables, which makes Eq.~\eqref{eq:discrepancy_model} true in a distributional sense.
The simple relationship \eqref{eq:discrepancy_model} directly impacts both the fitting protocol and error propagation to observables.
For example, if one assumes that $\delta\genobsth = 0$ and $\delta\genobsexp$ is normally distributed, this leads to a least squares log likelihood function for $\lecs$ given $\genobsexp$~\cite{WesolowskiExploringBayesianparameter2019}.
Here we use general properties of EFTs to develop a Bayesian model for $\delta\genobsth(\kinparvec)$ with a non-trivial correlation structure depending on $\kinparvec$.
Unless otherwise stated, we assume that the EFT has already been fit to experimental data and return to discuss the fitting procedure in Sec.~\ref{sub:application_types}.

Denote an observable $\genobs$ calculated with an $n$\textsuperscript{th}-order EFT ($\NkLO{n}$) as $\genobs_n$.
Suppose that the EFT expansion has only been computed up to $k$\textsuperscript{th} order.
Though $\genobs_k$ may work well as an approximation to reality, we seek a probability distribution for the full summation\footnote{
We are aware that EFTs are often \emph{asymptotic} in nature.
We seek the best possible estimate up to where the sum starts diverging.
Our assumption is that the divergence occurs sufficiently beyond the $k$th order that the truncation model error estimate up to order $\infty$ is a good approximation.
} $\genobs \equiv \genobs_\infty$ based on the convergence pattern of all known predictions $\genobs_0, \genobs_1, \dots, \genobs_k$.
That is, $\genobsth = \genobs_k$ and we wish to quantify $\delta\genobsth = \delta\genobs_k$.
This hierarchy of predictions could instead be written as a leading-order calculation $\genobs_0$ and a set of higher-order corrections $\Delta\genobs_n$, with $n\in \{1, 2, \cdots, k\}$.
Given this change of variables, the $n \leq k$ order prediction is
\begin{align} \label{eq:orders}
    \genobs_n(\kinparvec) = \genobs_0(\kinparvec) + \Delta\genobs_1(\kinparvec) + \cdots + \Delta\genobs_n(\kinparvec),
\end{align}
where each term in the series is known.

Without further knowledge of the convergence pattern in Eq.~\eqref{eq:orders}, it is difficult to make progress in approximating $\genobs$.
To proceed, we employ prior information about the construction of EFTs, that is, if the EFT is working as advertised, each correction should be roughly suppressed by the dimensionless expansion parameter $Q$---in accordance with the EFT power counting.
Here we assume for simplicity the following power counting: the $\NkLO{n}$ correction $\Delta\genobs_n$ is suppressed by $Q^n$, though other prescriptions, e.g., $Q^{2n}$, can be easily incorporated into this framework.
That is, up to dimensionful scales, one would generally expect $\genobs_0$ is $\order{Q^0}$, and $\Delta\genobs_n$ is $\order{Q^n}$.

Inspired by Eq.~\eqref{eq:orders} and the power counting of the EFT, Ref.~\cite{furnstahl_quantifying_2015} proposed the following factorization% of Eq.~\eqref{eq:orders}
\begin{align} \label{eq:Xk}
    \genobs_k(\kinparvec) = \genobsref(\kinparvec) \sum_{n=0}^k c_n(\kinparvec) Q^n(\kinparvec).
\end{align}
Here, $\genobsref$ is a dimensionful quantity that sets the scale of variation, while the $c_n$ are dimensionless observable coefficients.
Since all scales have been factored into $\genobsref$ and $Q$, the $c_n$ should be \emph{natural}, or order 1, assuming they have not been fine tuned.
Combinatorial factors, if known, should also appear in Eq.~\eqref{eq:Xk}, else the naturalness assumption may need to be modified to account for them.

\begin{figure*}[tb]
\captionsetup[subfloat]{captionskip=-15pt, margin=0pt}
\subfloat[]{%
\includegraphics{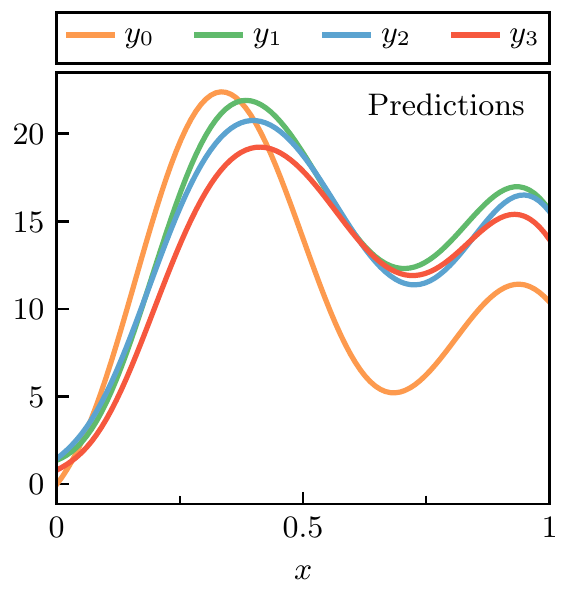}
\label{fig:pred_to_coeffs_preds}
}
\subfloat[]{%
\includegraphics{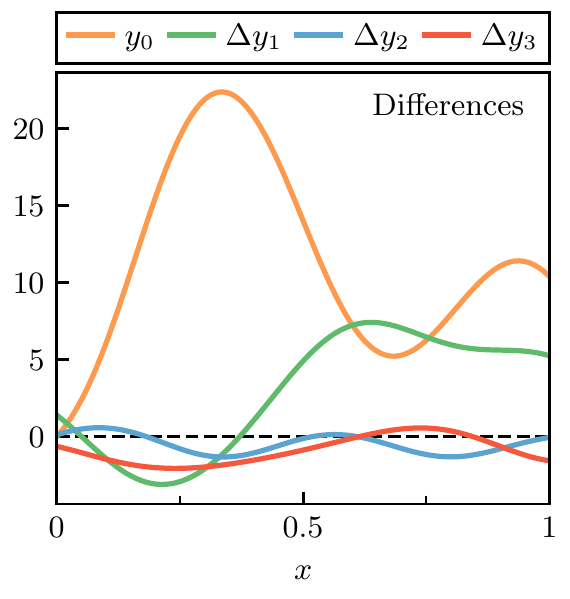}
\label{fig:pred_to_coeffs_diffs}
}
\subfloat[]{%
\includegraphics{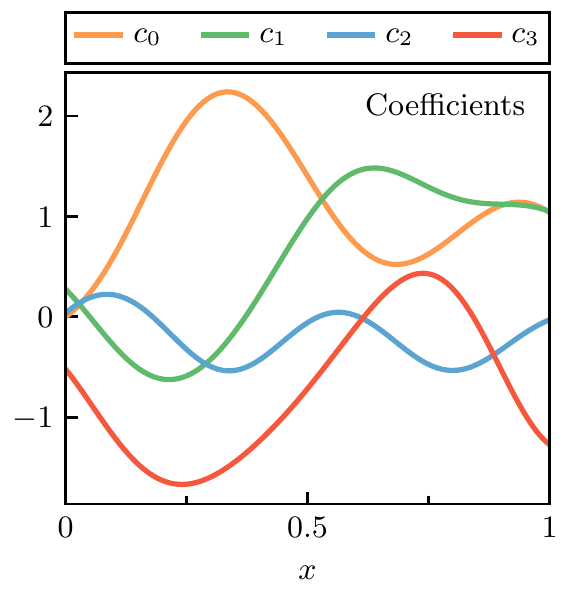}
\label{fig:pred_to_coeffs_coeffs}
}

\caption{The roadmap from EFT predictions to prediction coefficients: (a) the predictions themselves as a function of a generic variable $x$, (b) the leading-order prediction $\genobs_0$ and the increasingly suppressed 
order-by-order corrections $\Delta\genobs_n$, and (c) the dimensionless coefficients $c_n$ (using $\genobsref=10$ and $Q = 0.5$).
One might object that $|\Delta\genobs_3|$ is often larger than $|\Delta\genobs_2|$, or that the coefficients in (c) do not appear sufficiently random.
This is a consequence of a small sample size and seeing patterns in randomness: the coefficients $c_n$ shown above are actual random draws from an underlying GP and were used to build parts (a) and (b).
By chance $c_2$ is relatively small, and thus the correction $|\Delta\genobs_2|$ is actually smaller than $|\Delta\genobs_3|$ for this choice of $Q$.
It pays to remember this example when considering real EFT predictions.
}
\label{fig:preds_to_coeffs}
\end{figure*}

The road-map from continuous observable quantities $\genobs_0, \dots, \genobs_k$ to coefficients is shown visually for toy EFT predictions in Fig.~\ref{fig:preds_to_coeffs}.
Predictions for the four lowest orders of the EFT are plotted in Fig.~\ref{fig:preds_to_coeffs}(a), and show a steady convergence towards the final result.
Suppose that the \NkLO{3} EFT is the state of the art; thus our task is to estimate the uncertainty in $\genobs_3$.
The change of variables described by Eq.~\eqref{eq:orders} is illustrated in Fig.~\ref{fig:preds_to_coeffs}(b).
Note that there is a clear hierarchy to the corrections: they tend to become smaller as order increases.
Assume that $\genobsref$ is known, say, from dimensional analysis, and $Q$ is known from theory arguments.
The coefficients of the expansion can then be extracted using Eq.~\eqref{eq:Xk}, as shown in Fig.~\ref{fig:preds_to_coeffs}(c).
\emph{The hierarchy has vanished, and we are left with a set of curves that appear randomly drawn from an underlying process.}
Reference~\cite{MelendezBayesiantruncationerrors2017} displays $c_n$ for a collection of $NN$ scattering observables in chiral EFT, many of which exhibit these properties.

Before moving on, we would like to correct some common misconceptions about Eq.~\eqref{eq:Xk}.
We have made no assumptions about the structure of the expansion thus far.
In fact, Eq.~\eqref{eq:Xk} is \emph{not} a formal expansion of $\genobs$ in powers of $Q$.
Rather, we have only noted that if we assume values for $\genobsref$ and $Q$ along with the factorization of Eq.~\eqref{eq:Xk}, then there is a one-to-one correspondence between the predictions $\{\genobs_0,\dots,\genobs_k\}$ and the coefficients $c_n$ given by
\begin{align} \label{eq:constraints}
\begin{split}
    \genobs_0(\kinparvec) & \equiv \genobsref(\kinparvec) c_0(\kinparvec) \\ \Delta\genobs_n(\kinparvec) & \equiv \genobsref(\kinparvec) c_n(\kinparvec) Q^n(\kinparvec).
\end{split}
\end{align}
The coefficients are \emph{not} the LECs of the EFT\@.
Because we are making predictions that take the LECs as input, the $c_n$ are potentially complicated functions of the LECs, $Q$, and other variables.
Therefore, we do not have a general proof that the naturalness of the LECs propagates to the $c_n$; rather, it must be verified in each application.
If some of the $c_n$ do not conform to our assumptions, they may need to be left out of the analysis.

\begin{figure}[tb]
\includegraphics{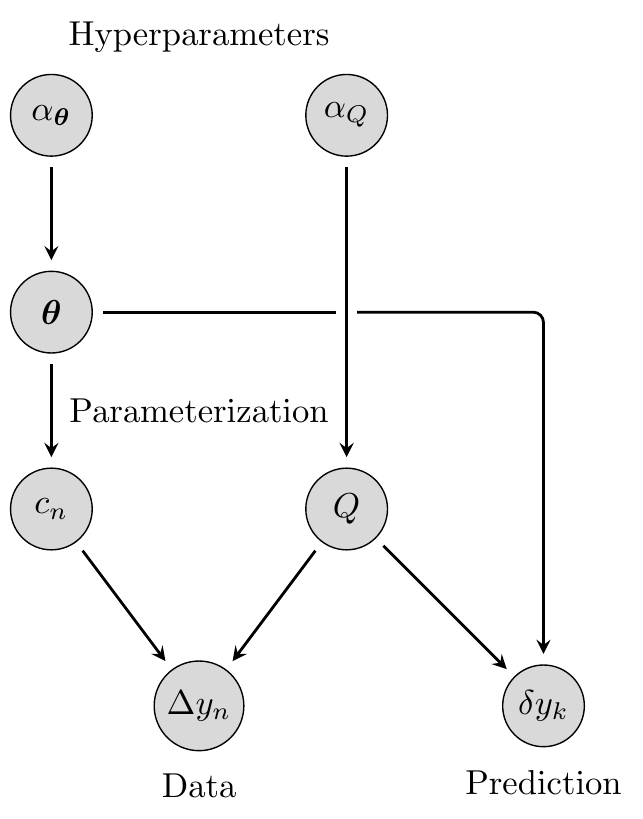}
\caption{A Bayesian network that demonstrates the causal connections between the random variables in our model.
Order-by-order corrections $\Delta\genobs_n$ are assumed to be composed of coefficients $c_n$, which are draws from a GP, and factors of $Q$, as in Eq.~\eqref{eq:constraints}.
The observed values of $\Delta\genobs_n$ are used to update our beliefs of the GP parameters $\param$ and the expansion parameter $Q$, which themselves are random variables governed by the hyperparameters $\alpha_{\param}$ and $\alpha_Q$, respectively.
For example, $\hypprior\hypm, \hypprior\hypdisp, \hypprior\hypdf, \hypprior\hyptau^2 \in \alpha_{\param}$ as assumed in Eq.~\eqref{eq:ninvchi2_prior}.
Once posteriors for $\param$ and $Q$ have been obtained, they can be used to create a posterior for $\delta\genobs_k$, hence quantifying the uncertainties in the predictions.
Note that information only flows from the differences $\Delta\genobs_n$ to the truncation error $\delta\genobs_k$ via the GP parameters $\param$ and the expansion parameter $Q$.
The reference scale $\genobsref$ is assumed to be known throughout and thus is not included in this diagram.}
\label{fig:bayesian_network_new}
\end{figure}

Our manipulations have thus far have simply defined our terminology, but the arguments above lead to a physics-based uncertainty model that can be tested against reality.
By a logical extension of Eq.~\eqref{eq:Xk}, the truncation error $\delta\genobs_k$ would consist of all terms not included in the sum, i.e.,
\begin{align} \label{eq:discrepancy_k_definition}
    \delta\genobs_k(\kinparvec) = \genobsref(\kinparvec) \sum_{n=k+1}^\infty c_n(\kinparvec) Q^n(\kinparvec).
\end{align}
We can now describe the details of the EFT convergence model.
By induction, we assume that the properties of the unobserved $c_n$ for $n > k$ are the same as the $c_n$ for $n\leq k$.
Specifically, we assume that the $c_n$ are independent and identically distributed (\iid) random curves.
Moreover, we have some prior knowledge of the curves' properties: they should be smooth and naturally sized, i.e., have a standard deviation of order 1.
Data from the lower-order predictions can then be used to refine the estimates for the sizes and shapes of all subsequent $c_n$ and, via Eq.~\eqref{eq:discrepancy_k_definition}, estimate the truncation error.
This is shown visually in the Bayesian network of Fig.~\ref{fig:bayesian_network_new}.

Because the coefficients are treated as random curves, the distribution from which they are drawn is known as a stochastic \emph{process}.
A stochastic process is essentially an infinite dimensional generalization of a probability distribution.
Processes associate a random variable with each point in a continuous domain, hence their infinite dimensionality, and correlations between the random variables at different points can be built in.
Through the appropriate assumption on this correlation structure, one can enforce desirable qualities in random draws from the process, such as continuity of the drawn functions and their derivatives.
GPs are particularly useful and will be employed throughout this paper.

%%%%%%%%%%%%%%%%%%%%%%%%%%%%%%%%%%%%%%%%%%%%%%%%%%%%%%%%%%%%%%%

\subsection{A Brief Introduction to Gaussian Processes}
\label{sub:gaussian_processes}

\subsubsection{Definitions}
\label{ssub:definitions}

GPs are a popular tool for nonparametric regression due to their flexible nature and tractable analytic forms.
They are used in fields such as statistics, machine learning, and geostatistics, where GP regression is known as \emph{kriging}~\cite{sacks1989design,cressie1992statistics,rasmussen_gaussian_2006}.
For appropriate priors, GPs arise as the limit of a specific type of neural network~\cite{neal_bayesian_1996} and are equivalent to common interpolants, e.g., splines~\cite{mackay_introduction_1998}.
Here we will review the most common applications of GPs: interpolation and regression.
Then we discuss the calibration of the GP parameters, which is a crucial step in our truncation error model.
For more in-depth discussions of GPs, see Refs.~\cite{rasmussen_gaussian_2006,mackay_introduction_1998,mackay_information_2003}.

The defining quality of a GP is~\cite{rasmussen_gaussian_2006}
\begin{quote}
A Gaussian process is a collection of random variables, any finite number of which have a joint Gaussian distribution.
\end{quote}
A GP is specified by a mean function $m(\kinparvec)$ and a positive semi-definite covariance function, or kernel, $\kernel(\kinparvec, \kinparvec')$, where $\kinparvec \in \mathbb{R}^d$.
Assuming both the mean and covariance functions are known, a GP $\func(\kinparvec)$ is denoted by\footnote{\label{statnote}The $z \sim \cdots$ notation is a common shorthand in statistical literature for ``$z$ is distributed as.'' Some authors use $\pr(z) = \cdots$ as well~\cite{GelmanBayesianDataAnalysis2013}. Also ``$z\given I$'' is read as ``$z$ given $I$.''}
\begin{align} \label{eq:gp_definition}
  f(\kinparvec) \sim \GP[m(\kinparvec), \kernel(\kinparvec, \kinparvec')].
\end{align}
We do not operate on the infinite-dimensional object in practice; rather, the definition of GPs allows us to compute using a finite number of points.
Let $\kinparvecset = \{\kinparvec_i\}_{i=1}^N$ be a set of $N$ input points and $\funcvec = \{\func(\kinparvec_i)\}_{i=1}^N$ be the corresponding set of function values.%
\footnote{To specify how the GP enters in our truncation model, we must work in multiple vector spaces. The bold notation for vectors signals an element of $\mathbb{R}^N$ while we use arrows to indicate a vector in the space of EFT orders $\mathbb{R}^{\nc}$.  Bold with arrows is used for elements of both spaces.  The $d$ dimensional input space is not explicitly indicated.  This notation is summarized in Table~\ref{tab:notation} in Appendix~\ref{app:derivations}.}
Henceforth, we refer to coefficient values computed at particular input points as ``data'', because these are the points used to train the GP emulator.

Define $\meanvec = m(\kinparvecset) \in \mathbb{R}^N$ and $K = \kernel(\kinparvecset, \kinparvecset) \in \mathbb{R}^{N\times N}$.
$\funcvec$ is then distributed as the multivariate normal
\begin{align} \label{eq:gaussian_dist}
  \funcvec \given \kinparvecset \sim \normal(\meanvec, K).
\end{align}
The fact that Eq.~\eqref{eq:gp_definition} implies Eq.~\eqref{eq:gaussian_dist} is the definition of a GP in a mathematical form.
Often the explicit conditioning on input points $\kinparvecset$ from $\funcvec \given \kinparvecset$ is dropped for notational simplicity.
The $N$-dimensional Gaussian distribution of Eq.~\eqref{eq:gaussian_dist} can then easily be used to draw samples of $\funcvec$ using the Cholesky decomposition of $K$.

The mean function is an \emph{a priori} ``best guess'' of $f$, and should capture overall trends in the data; the GP then soaks up deviations from this mean in accordance with the chosen kernel~\cite{GelmanBayesianDataAnalysis2013}.
% A linear model $m(\kinparvec) = \beta_0 + \kinparvec^\trans \beta$ or basis function model $m(\kinparvec) = \basisfunc(\kinparvec)^\trans \beta$, where $\beta$ are regression coefficients, are convenient choices due to their analytic tractability and flexibility.
Basis function models for $m(\kinparvec)$ are convenient choices due to their analytic tractability and flexibility~\cite{GelmanBayesianDataAnalysis2013}. 
But the effects of mean functions can also be  shuffled into the kernel (see Appendix~\ref{app:derivations}) or simply subtracted from the data in preprocessing, hence many authors set the mean to zero.
The convergence model proposed here uses a constant (possibly unknown) mean $\mu$, which we leave in the mean function slot for clarity.

\begin{figure}[tb]
\includegraphics{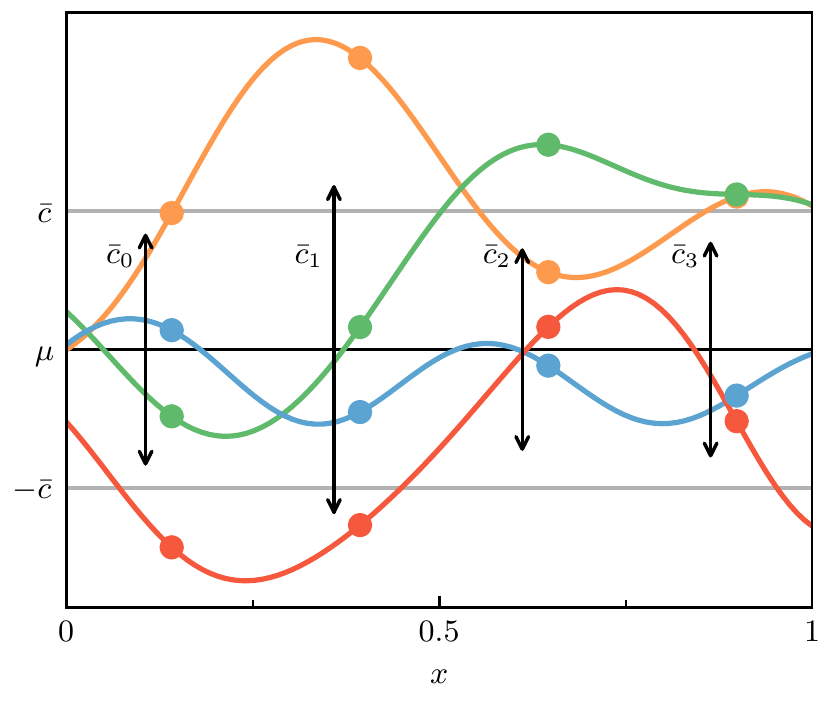}
\caption{The random draws from a GP used in Fig.~\protect\subref*{fig:pred_to_coeffs_coeffs}.
The GP has mean $\mu$, marginal standard deviation $\sdth$, and a Gaussian covariance function with length scale $\ell$, each of which are annotated on the plot.
The ``uncorrelated'' and ``fully correlated'' truncation error model discussed in Ref.~\cite{WesolowskiExploringBayesianparameter2019} correspond to this model with $\ell=0$ and $\ell=\infty$, respectively, and $\mu=0$.
By using a finite set of points along each curve, shown as dots, these parameters can be estimated (or, calibrated) rather than assumed.
This GP model is contrasted with the pointwise model from Refs.~\cite{furnstahl_quantifying_2015,MelendezBayesiantruncationerrors2017}, where data at each $\kinparvec_i$ are used to estimate a $\sdth_i \equiv \sdth(\kinparvec_i)$ (and where there is no $\ell$).
}
\label{fig:random_annotated_curves}
\end{figure}

One of the most popular covariance functions is the squared exponential (a.k.a.\ radial basis function or Gaussian).
It is defined by
\begin{align}
  \kernel(\kinparvec, \kinparvec';\sdth, L) & = \sdth^2 r(\kinparvec, \kinparvec';L) \notag \\
  & = \sdth^2 e^{-\frac{1}{2}(\kinparvec-\kinparvec')^\trans L^{-1}(\kinparvec-\kinparvec')}, \\
  & \to \sdth^2 e^{-(\kinparvec-\kinparvec')^\trans (\kinparvec-\kinparvec') / 2\ell}, \label{eq:se_kernel}
\end{align}
where $r(\kinparvec, \kinparvec';L)$ is the correlation function
and $L$ is a $d \times d$ diagonal matrix of correlation length parameters, allowing one to model
cases where the correlation length varies by direction~\cite{rasmussen_gaussian_2006}.
% anisotropic responses, i.e., cases where the correlation length parameter differs in different dimensions of the response.~\cite{rasmussen_gaussian_2006}.
For simplicity our examples are in one dimension ($d=1$), so $L$ reduces to a single correlation length $\ell$.
%For simplicity our examples are in one dimension, where $L=\ell\, \identity = \ell$.
%For simplicity our examples are in one dimension, and $L$ is a $d\times d$ diagonal matrix of correlation length parameters, allowing one to model anisotropic responses, i.e., cases where the correlation length parameter differs in different dimensions of the response. Here $L = \ell \identity$, where the identity is one-dimensional.
% The assumption that $L \propto \identity$ is likely inadequate for higher-dimensional problems, but $L$ can still be estimated and model-checked using the methods presented here.
%which makes the kernel \emph{isotropic}.
Functions drawn from a GP with a squared exponential kernel are infinitely differentiable.
The marginal variance $\sdth^2$ controls the width of variation about the mean, while the length scale $\ell$ controls the correlation between points in the input domain, see Fig.~\ref{fig:random_annotated_curves}.
Some have argued that the squared exponential is \emph{too} smooth, and thus not applicable in many situations~\cite{SteinInterpolationSpatialData1999}.
Thus, an alternative kernel is the Mat\'ern, which provides an extra parameter $\nu$ and results in realizations that are $\lceil\nu\rceil - 1$ times differentiable, reproducing the squared exponential in the limit $\nu \to \infty$.

The squared exponential kernel, along with the Mat\'ern, are \emph{stationary}, which means that they only depend on the absolute difference in input space $|\kinparvec - \kinparvec'|$.
Draws from a stationary process would, on average, behave similarly across the domain of interest.
Stationarity is a strong assumption about the physics of the system, which can be advantageous in data-sparse regimes that could benefit from extra constraints.
But its validity should be checked.
We introduce diagnostics that can point to issues with the assumption of stationarity, along with the choices of mean and covariance, in Sec.~\ref{sec:model_checking}.

We assume the functional form of the mean and covariance functions (e.g., squared exponential) are known. 
Their parameters,\footnote{We choose to call $\mu$, $\sdth$, etc., the \emph{parameters} of the GP to distinguish them from the hyperparameters we will introduce later. They are more regularly referred to as hyperparameters.} on the other hand, can be updated based on observed data.
We denote the set of all parameters in both $m$ and $\kernel$ (e.g., $\mu$, $\sdth^2$, $\ell$) as $\param$.
We are careful to write $\cdot \given \param$ when $\param$ is given (i.e., \emph{conditioned} upon) and omit $\param$ otherwise.

\begin{figure}[tb]
\includegraphics{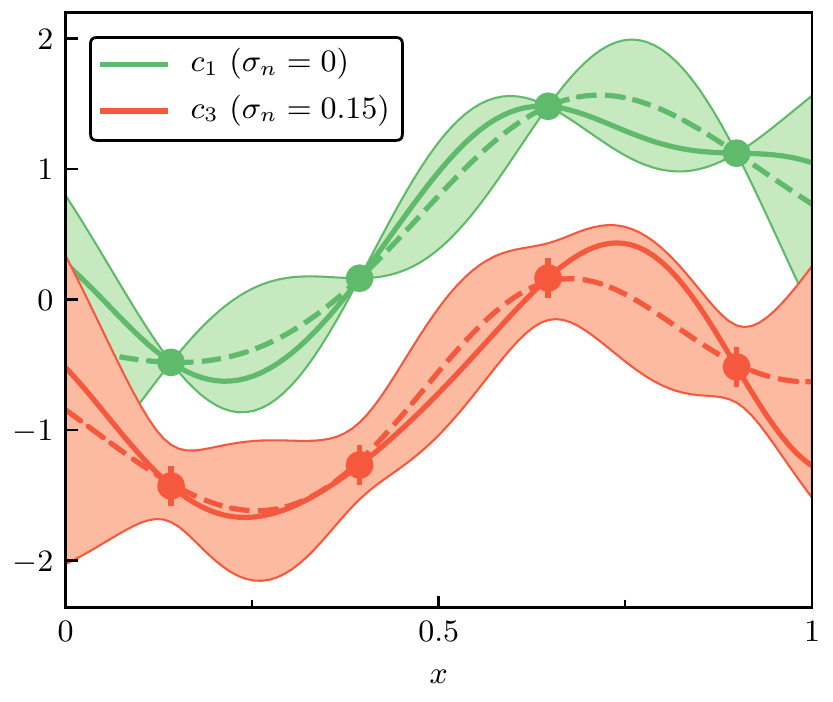}
\caption{Examples of interpolation ($\sdnugget = 0$) and regression ($\sdnugget = 0.15$) using the GP described in Fig.~\ref{fig:random_annotated_curves}.
The true curves are shown as solid lines and the interpolants are shown as dashed lines.
The bands denote $2\sigma$ marginal confidence intervals.
As one might expect, there is a bubble-like structure to the interpolation error when between known points.
The process will eventually return to the mean when far from the support of the data.
}
\label{fig:gp_interp_and_noise_example}
\end{figure}

\subsubsection{Interpolation and Regression}
\label{ssub:interpolation_and_regression}

To use GPs for interpolation or regression, we must be able to incorporate (noisy) observations into our predictions.
Interpolation, in this case, simply refers to the procedure of finding a curve that passes \emph{directly} through known data, regardless of whether the predictions are made within or outside the support of the data.
Consider the partition $\kinparvecset = \transpose{\begin{bmatrix} \kinparvecset_1 & \kinparvecset_2 \end{bmatrix}}$ and $\funcvec = \func(\kinparvecset) = \transpose{\begin{bmatrix} \funcvec_1 & \funcvec_2 \end{bmatrix}}$ into $N_1$ training and $N_2$ test points, with corresponding partitioning of $\meanvec$ and $K$.
We would like to predict the unseen function values $\funcvec_2$ given observed function values $\funcvec_1$.
By definition of GPs, their $N_1 + N_2$ dimensional joint distribution is Gaussian:
\begin{align} \label{eq:GP_joint_dist}
  \begin{bmatrix}
    \funcvec_1 \\
    \funcvec_2
  \end{bmatrix}
  \Big|\, \kinparvecset, \param
  \sim
  \normal{\left(
  \begin{bmatrix}
    \meanvec_1 \\ \meanvec_2
  \end{bmatrix}
  ,
  \begin{bmatrix}
    K_{11} & K_{12} \\
    K_{21} & K_{22}
  \end{bmatrix}
  \right)}.
\end{align}
Then by standard Gaussian identities~\cite{rasmussen_gaussian_2006},
\begin{align} \label{eq:gp_conditional_pred}
  \funcvec_2 \given \kinparvecset, \funcvec_1, \param \sim \normal(\tilde\meanvec_2, \tilde K_{22}),
\end{align}
where
\begin{align}
  \tilde\meanvec_2 & \equiv \meanvec_2 + K_{21} K_{11}^{-1} (\funcvec_1 - \meanvec_1) \label{eq:interp_mean} \\
  \tilde K_{22} & \equiv K_{22} - K_{21} K_{11}^{-1} K_{12}. \label{eq:interp_cov}
\end{align}
Since we assume that $K$ can factorize into a marginal variance $\sdth^2$ and a
correlation matrix $R$, i.e., $K = \sdth^2 R$, 
this can be written as
\begin{align}
  \tilde\meanvec_2 & = \meanvec_2 + R_{21} R_{11}^{-1} (\funcvec_1 - \meanvec_1) \\
  \tilde K_{22} & = \sdth^2\left(R_{22} - R_{21} R_{11}^{-1} R_{12}\right) \equiv \sdth^2 \tilde R_{22}.
\end{align}
This work considers a hierarchy of noiseless computer simulations as its data, so interpolation is the primary fitting method and an example is shown in Fig.~\ref{fig:gp_interp_and_noise_example}.
Nevertheless, interpolation may require a small amount of noise called a \emph{nugget} for the numerical stability of the matrix inversion~\cite{Ababouconditionnumbercovariance1994,AndrianakiseffectnuggetGaussian2012,Mohammadianalyticcomparisonregularization2016,Pepelyshevrolenuggetterm2010,RanjanComputationallyStableApproach2011}.

To make predictions of $\funcvec_2$ assuming Gaussian white noise in the observed data $\funcvec_1$, one need only make the replacement $K_{11} \to K_{11} + \sdnugget^2\identity_{N_1}$ in Eqs.~\eqref{eq:interp_mean} and~\eqref{eq:interp_cov}, leaving $K_{21}$, etc., unchanged.
This is what we refer to as regression.
The identity matrix is denoted by $\identity$ with the rank as subscript, and here $\sdnugget^2$ is the nugget used to regularize the matrix inversion.
To make predictions that also include the effects of noise, one must also make the replacement $K_{22} \to K_{22} + \sdnugget^2\identity_{N_2}$.
An example of regression with noise is shown in Fig.~\ref{fig:gp_interp_and_noise_example}.
We include the nugget via $\sdth^2(R + \sdnugget^2\identity_N)$ to integrate $\sdth^2$ out analytically.

\subsubsection{Calibration}
\label{ssub:calibration}

Until now, we have assumed that the GP parameters $\param$ are known and correct, but that is rarely ever the case in practice.
We often want to find the $\param$ that produce the best fit in interpolation/regression, or $\param$ may be of interest in its own right.
The procedure of finding the posterior for $\param$ values is known as \emph{calibration}, or simply Bayesian inference for the model parameters~\cite{SantnerDesignAnalysisComputer2018}.

Given data $\funcvec$ at input points $\kinparvecset$, the posterior for $\param$ is
\begin{align} \label{eq:param_post}
  \pr(\param \given \kinparvecset, \funcvec) \propto \pr(\funcvec \given \kinparvecset, \param) \pr(\param).
\end{align}
If $\func$ is modeled as a GP, then Eq.~\eqref{eq:param_post} is, up to a normalization constant, Eq.~\eqref{eq:gaussian_dist} multiplied by the priors.
The posterior can be sampled via Markov chain Monte Carlo (MCMC) methods, or optimized using gradient descent to find its maximum \emph{a posteriori} (MAP) value $\param_{\mathrm{MAP}}$.
When a single value for $\param$ is used as a ``best guess,'' e.g., $\param_{\mathrm{MAP}}$, this is called a \emph{point estimate}.

For certain cases, one can find analytic posteriors for parameters in $\param$ by making use of \emph{conjugate priors}, whose posteriors have the same functional form as the prior~\cite{GelmanBayesianDataAnalysis2013}.
For our applications only a few features of the prior information are important; we are generally insensitive to details of the priors' functional form. Our strategy is then to make use of conjugate priors where they are flexible enough to capture distributions' relevant features, and find MAP values for the remaining parameters. 

Our conjugate priors on the GP parameters $\mu$ and $\sdth^2$ are then characterized by hyperparameters that quantify
the mean and variance for the prior distributions of both, as well as the heaviness of the $\sdth^2$ distribution's tail. Our model is also easily built into existing MCMC samplers, such as PyMC3~\cite{salvatier_probabilistic_2016}, where any functional form for the priors can be used.

%For our applications, we are generally insensitive to the precise functional form of the priors while the relevant conjugate priors are flexible enough to capture the important features of the prior information. 
%In this work we make use of conjugate priors where possible and justified, and find MAP values for the remaining parameters.

%conjugate priors are unnecessary.

To make predictions unconditional on $\param$, which take all plausible values into account, one must compute
\begin{align} \label{eq:f_unconditional_prediction}
  \pr(\funcvec_2 \given \kinparvecset, \funcvec_1) = \int \pr(\funcvec_2 \given \kinparvecset, \funcvec_1, \param) \pr(\param \given \kinparvecset_1, \funcvec_1) \dd{\param}.
\end{align}
This can be approximated 
%by replacing integrals by $\param_{\mathrm{MAP}}$, 
by replacing the integral over $\param$ by evaluation of the integrand at $\param_{\mathrm{MAP}}$,
computed numerically with Monte Carlo methods, or evaluated analytically with the help of conjugacy.
For example, Appendix~\ref{app:derivations} shows how $\mu$ and $\sdth^2$ are integrated out of Eq.~\eqref{eq:f_unconditional_prediction} analytically via conjugacy.

This brief introduction to GPs is enough to set the stage for our model of truncation errors.

%%%%%%%%%%%%%%%%%%%%%%%%%%%%%%%%%%%%%%%%%%%%%%%%%%%%%%%%%%%%%%%

\subsection{The GP Model of EFT Convergence}
\label{sub:gp_truncation}

Recall from Sec.~\ref{sub:setup} that after making the transformation from prediction functions $\{\genobs_n\}$ to coefficient functions $\{c_n\}$, the coefficients appeared as naturally sized random curves.
Here we will specify exactly what we mean by random and discuss how we can extract meaningful insight from this assumption.

We formalize our EFT convergence assumptions via Eqs.~\eqref{eq:Xk} and~\eqref{eq:discrepancy_k_definition}, where the coefficients $c_n$ are independent draws from a single underlying GP.
We assume a constant mean function $m(\kinparvec) = \mu$ for simplicity and a kernel $\kernel(\kinparvec, \kinparvec';\sdth, \ell) = \sdth^2 r(\kinparvec,\kinparvec';\ell)$ that need not be stationary.
Thus,
\begin{align}
    c_n(\kinparvec) \given \param & \overset{\text{\tiny iid}}{\sim} \GP[\mu, \sdth^2r(\kinparvec,\kinparvec';\ell)], \label{eq:cn_iid} \\
  \param & \equiv \{\mu, \sdth^2, \ell\}. \label{eq:param}
\end{align}
Figure~\ref{fig:random_annotated_curves} shows random draws from such a process and a comparison to the model described in Refs.~\cite{furnstahl_quantifying_2015,MelendezBayesiantruncationerrors2017}.

The parametrization of Eq.~\eqref{eq:cn_iid} can be generalized as appropriate.
Non-constant mean functions can be incorporated by promoting $\mu$ to a vector of regression coefficients multiplying basis functions.
For the model of the $c_n$'s that we use here this basis would contain only one, constant, function. But if one wants to build specific structure into the mean function for $\genobs_k$ and $\delta\genobs_k$, then the notion of a basis is useful for discussing the distribution of these quantities. We discuss that more general case, adopting vector and matrix notation, in Appendix~\ref{app:derivations}. 
% Non-constant mean functions can be used by promoting $\mu$ to a length $p$ vector of regression coefficients multiplying basis functions $\basisfunc(\kinparvec)$, i.e., $m(\kinparvec) = \transpose\basisfunc(\kinparvec)\mu$.
% Though for the $c_n$ we assume that $\basisfunc(\kinparvec) = 1$, the notion of a basis is useful for discussing the distribution for $\genobs_k$ and $\delta\genobs_k$.
% We will use matrix notation, e.g., $\transpose\basisfunc(\kinparvec)\mu$, despite these quantities being scalars in our case, for generality.
The factorization of the kernel into a marginal variance $\sdth^2$ and a correlation function $r$ that depends on a correlation length $\ell$ may seem restrictive.
On the contrary, this kernel can incorporate a wide range of functional structures since $r$ can be non-stationary in general and $\ell$ can stand in for any set of correlation parameters.
Thus here we make $\mu$, $\sdth^2$, and $\ell$ the explicit parameters of our GP model: this allows us to show how they can be calibrated or integrated out of this system.

The distribution of $\delta\genobs_k$, and thus the full prediction $\genobs = \genobs_k + \delta\genobs_k$, can be derived using the remarkable property of Gaussian random variables: they are closed under addition and matrix multiplication.
Let $X\sim\normal(\mu_1, K_1)$ and $Y \sim \normal(\mu_2, K_2)$ be length-$N$ independent random variables and let $A,B$ be known $M\times N$ matrices.
Then
\begin{align} \label{eq:normal_sum}
  AX+BY \sim \normal(A\mu_1 + B\mu_2, A K_1\transpose{A} + B K_2\transpose{B}).
\end{align}
Equation~\eqref{eq:discrepancy_k_definition}, which defines the truncation error $\delta\genobs_k$, is a geometric sum over normally distributed variables.
Using Eqs.~\eqref{eq:cn_iid} and~\eqref{eq:normal_sum}, one can readily find the prior
\begin{align} \label{eq:discr_k_prior}
    \delta\genobs_k(\kinparvec) \given \param, Q \sim \GP[\discrmean{k}(\kinparvec), \sdth^2\discrcorr{k}(\kinparvec, \kinparvec'; \ell)],
\end{align}
where
\begin{align}
    \discrmean{k}(\kinparvec) & \equiv  
     \genobsref(\kinparvec)\frac{Q(\kinparvec)^{k+1}}{1 - Q(\kinparvec)}\mu 
     \equiv \discrbasisfunctrans{k}(\kinparvec)\mu  \label{eq:truncation_basisfunc} \\
    \discrcorr{k}(\kinparvec, \kinparvec';\ell) & \equiv \genobsref(\kinparvec)\genobsref(\kinparvec')\frac{[Q(\kinparvec)Q(\kinparvec')]^{k+1}}{1 - Q(\kinparvec)Q(\kinparvec')} r(\kinparvec, \kinparvec';\ell). \label{eq:truncation_corrfunc}
\end{align}
We have defined the basis function $\discrbasisfunctrans{k}(\kinparvec)$ as it will recur below.
Although $\discrcorr{k}$ may look like a correlation matrix, the inclusion of the $\genobsref$ and $Q$ factors mean that $\discrcorr{k}(\kinparvec, \kinparvec;\ell) \neq 1$, in general.
The marginal variance of Eq.~\eqref{eq:discr_k_prior} is thus, in general, $\kinparvec$ dependent and equal to $\sdth^2\discrcorr{k}(\kinparvec, \kinparvec; \ell)$.

Equations~\eqref{eq:discr_k_prior}--\eqref{eq:truncation_corrfunc} define the functional form of the truncation errors, but how should one estimate their parameters in a statistically rigorous manner?
The simplest (and least rigorous) way is to assume some reasonable values for $\param$ and $Q$ based on expert knowledge of the system.
For example, $\mu=0$ and $\sdth \approx 1$ is a good place to start if one can assume that the coefficients are naturally sized and equally likely to be positive or negative.
Then $\ell$ and $Q$ could be based on arguments specific to the observable and EFT under consideration.
Again, $\genobsref$ is assumed to be known in this work, possibly based on dimensional analysis for the observable.
The next best way would be to combine expert knowledge of the system with the known low-order predictions to compute point estimates of $\param$ and $Q$.
(Remember, information only propagates from the order-by-order predictions to $\delta\genobs_k$ via $\param$ and $Q$, as shown in Fig.~\ref{fig:bayesian_network_new}.)
Finally, the most rigorous Bayesian method is to compute the full posterior for $\param$ and $Q$ and marginalize (integrate) over all of their possible values.
Depending on the specific application of this model and the needs of the user, different degrees of rigor may be warranted.

\begin{figure}[tb]
\includegraphics{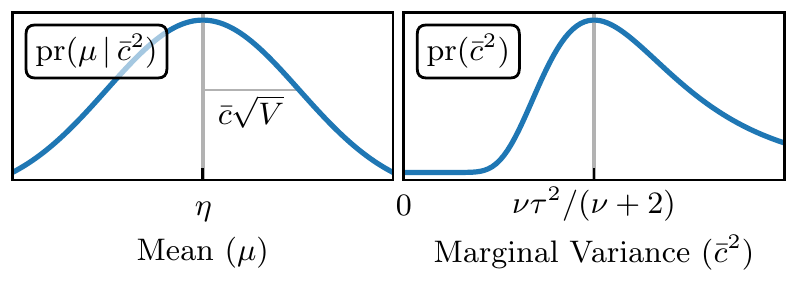}
\caption{Slices of the distribution $\mu,\sdth^2 \sim \ninvchisq(\hypm, \hypdisp, \hypdf, \hyptau)$.
The left plot is a Gaussian while the right is a scaled inverse $\chi^2$ distribution.}
\label{fig:priors}
\end{figure}

Assuming that we want to combine expert knowledge of the system with the order-by-order EFT predictions to estimate $\delta\genobs_k$, we must start by formalizing our knowledge as Bayesian priors.
We have checked the prior dependence of our pointwise model in previous works and found that truncation error estimates are fairly insensitive to their exact choice~\cite{furnstahl_quantifying_2015,MelendezBayesiantruncationerrors2017}.
Therefore, we choose to work with conjugate priors, which yield intuitive analytic posteriors and point estimates for the GP parameters $\param$.
We place a (scaled) normal-inverse-$\chi^2$ prior\footnote{
    A more common and mathematically equivalent choice is the normal-inverse-gamma prior.
    We find the normal-inverse-$\chi^2$ prior more intuitive to work with in this context and thus promote its use.
} on $\mu$ and $\sdth^2$,
\begin{align} \label{eq:ninvchi2_prior}
  \mu,\sdth^2 \sim \ninvchisq(\hypprior\hypm, \hypprior\hypdisp, \hypprior\hypdf, \hypprior\hyptau^2).
\end{align}
This joint prior factorizes as $\pr(\mu, \sdth^2) = \pr(\mu \given \sdth^2) \pr(\sdth^2)$, where
\begin{align}
  \mu \given \sdth^2 & \sim \normal(\hypprior\hypm, \sdth^2 \hypprior\hypdisp) \label{eq:mu_given_sigma_prior} \\
  \sdth^2 & \sim \invchisq(\hypprior\hypdf, \hypprior\hyptau^2). \label{eq:sigma_prior}
\end{align}
The inverse $\chi^2$ density is given by
\begin{align} \label{eq:invchi2_density}
  f(z;\hypdf, \hyptau^2) = \frac{(\hypdf\hyptau^2/2)^{\hypdf/2}}{\Gamma(\hypdf/2) z^{1+\hypdf/2}} \exp(-\frac{\hypdf\hyptau^2}{2z}).
\end{align}
See Fig.~\ref{fig:priors} for a graphical representation of Eqs.~\eqref{eq:mu_given_sigma_prior}--\eqref{eq:sigma_prior}.
The meaning of the hyperparameters $\hypprior\hypm, \hypprior\hypdisp, \hypprior\hypdf, \hypprior\hyptau^2$ becomes clear from Eqs.~\eqref{eq:mu_given_sigma_prior}--\eqref{eq:sigma_prior}.
\begin{itemize}
    \item \emph{The prior mean $\hypprior\hypm$}: Our best guess for $\mu$ before seeing the data.
    \item \emph{The prior dispersion\footnote{We have not found a standard name for this quantity in the literature. The dispersion, a generic term describing the variability or spread of a distribution, is as good as any.} $\hypprior\hypdisp$}: When combined with $\sdth^2$, this is our estimated uncertainty in $\mu$. It makes sense that if the spread of the $c_n$ is large, then we would be less certain about their mean.
    \item \emph{The prior degrees-of-freedom $\hypprior\hypdf$}: This quantity essentially describes our uncertainty in $\sdth^2$.
    For $\hypdf=0$, Eq.~\eqref{eq:invchi2_density} reduces to the scale invariant Jeffreys prior used in Ref.~\cite{MelendezBayesiantruncationerrors2017}, whereas for $\hypdf\to\infty$, it becomes sharply peaked.
    \item \emph{The prior scale $\hypprior\hyptau^2$}: For large $\hypdf$, this is our best guess for $\sdth^2$. The mode and mean of Eq.~\eqref{eq:invchi2_density} are given by $\hypdf\hyptau^2/(\hypdf\pm2)$, respectively.
    The mean is only defined for $\hypdf>2$.
\end{itemize}
There are no conjugate priors for $\ell$ or $Q$.
Reasonable choices could be a strictly positive log-normal prior for $\ell$ and a beta prior for $Q$, since it is restricted to be between zero and one.

Once priors have been chosen for $\param$ and $Q$, the rest of the process is algorithmic.
Given order-by-order data $\genobsvecset \equiv \{\genobsset_i\}_{i\leq k}$, one can derive analytic posteriors for all of $\param$ and $Q$.
See Appendix~\ref{app:derivations} for details on how this is done. From now on we assume that $\ell$ and $Q$ are known \emph{a priori} or point estimates are obtained from their posteriors.
Analytic results for the truncation error posterior follow.
Formulae in which $\ell$ and $Q$ are marginalized over are more complicated and not analytic.
Point estimates of $\ell$ and $Q$ are often good approximations in our application, making numerical integration a needless complication. MCMC sampling can be used in cases where this approximation is not adequate.

Conjugacy ensures that the posterior for $\mu, \sdth^2$ has the same functional form as the prior.
That is,
\begin{align} \label{eq:mu_sigma_posterior}
  \mu, \sdth^2 \given \genobsvecset, \ell, Q \sim \ninvchisq(\cond\hypm, \cond\hypdisp, \cond\hypdf, \cond\hyptau^2)
\end{align}
for updated hyperparameters $\cond\hypm$, $\cond\hypdisp$, $\cond\hypdf$, and $\cond\hyptau^2$ given by Eqs.~\eqref{eq:hypm_update_corr}--\eqref{eq:hyptau_update_corr}.
The path forward depends on whether point estimates of $\mu$ and $\sdth^2$ are sufficient.
If so, then the mean values $\E[\cdot]$ of $\mu$ and $\sdth^2$, given by
\begin{align}
    \E[\mu \given \genobsvecset, \ell, Q] & = \cond\hypm \label{eq:mean_point_estimate} \\
    \E[\sdth^2 \given \genobsvecset, \ell, Q] & = \frac{\cond\hypdf\cond\hyptau^2}{\cond\hypdf-2} \label{eq:sdth_point_estimate}
\end{align}
can be used as point estimates in Eq.~\eqref{eq:discr_k_prior}.
It is interesting to note that these estimates are Bayesian analogs to standard frequentist estimators for the mean and variance.
If one integrates over all possible values of $\mu$ and $\sdth^2$, then the posterior is a \emph{Student-$t$ process} (TP)~\cite{ShahStudenttProcessesAlternatives2014},
\begin{align} \label{eq:discr_k_posterior_student_t}
  \delta\genobs_k & (\kinparvec) \given \genobsvecset, \ell, Q \sim \\ & \TP_{\cond\hypdf}\!{\left\{\discrbasisfunctrans{k}(\kinparvec)\cond\hypm, \cond\hyptau^2 \left[\discrcorr{k}(\kinparvec,\kinparvec'; \ell) + \discrbasisfunctrans{k}(\kinparvec)\cond\hypdisp\discrbasisfunc{k}(\kinparvec')\right]\right\}}. \notag
\end{align}

Like a GP, any finite collection of random variables from a TP has a joint Student-$t$ distribution.
Equation~\eqref{eq:discr_k_posterior_student_t} is centered at our best guess for $\discrbasisfunctrans{k}(\kinparvec)\mu$ but also includes the uncertainty in $\mu$ via $\discrbasisfunctrans{k}(\kinparvec)\cond\hypdisp\discrbasisfunc{k}(\kinparvec')$ and the uncertainty in $\sdth^2$ via the heavy tails of the $t$ distribution (which are heavier for smaller $\cond\hypdf$).
The $\kappa(\kinparvec,\kinparvec')$ in $\TP_\hypdf[m(\kinparvec), \kappa(\kinparvec, \kinparvec')]$ is not the covariance function in our notation, but instead a scale function.
The covariance between the response at $x$ and $x'$ for a TP is given by $\cond\hypdf \kappa(\kinparvec,\kinparvec')/(\cond\hypdf-2)$ and is only defined for $\cond\hypdf > 2$.
Given this fact, note the relationship between the point estimate in Eq.~\eqref{eq:sdth_point_estimate} and the covariance in Eq.~\eqref{eq:discr_k_posterior_student_t}.

%%%%%%%%%%%%%%%%%%%%%%%%%%%%%%%%%%%%%%%%%

\subsection{Application Types}
\label{sub:application_types}

We have shown how to arrive at a physically motivated distribution for the truncation error, which we will now use as a springboard to discuss how our EFT convergence model is applicable to two situations that can arise when fitting and predicting.
\begin{enumerate}
    \item
    For certain physical systems, predictions may be expensive to compute.
    This can affect the fitting of the LECs, where predictions must be made repeatedly to find a best fit (or posterior), but can also affect predictions with fixed LECs for particularly expensive systems.
    For example, fitting chiral EFT beyond the two-body sector becomes computationally expensive.
    \item Certain observables may have symmetry constraints, which provides information about contributions from higher orders in the EFT\@.
    For example, certain spin observables in the $NN$ sector of chiral EFT must vanish at $\theta=0^\degr$ or $180^\degr$.
    Thus, the distribution for the truncation error should reflect this fact.
\end{enumerate}
Here we discuss how our correlated convergence model can solve both of these problems, and how to use this model when fitting an EFT\@.
We assume that point estimates from Eqs.~\eqref{eq:mean_point_estimate} and~\eqref{eq:sdth_point_estimate} are used so that everything remains Gaussian, and leave the TP case for Appendix~\ref{app:derivations}.

\begin{figure*}[tb]
\captionsetup[subfloat]{captionskip=-147pt, margin=4pt}
\subfloat[]{%
\includegraphics{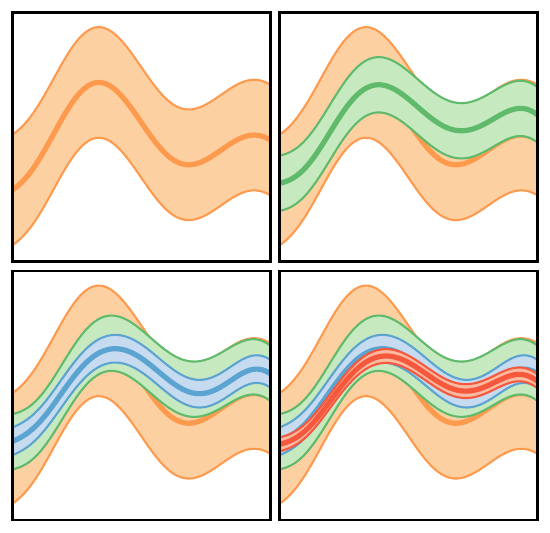}
\label{fig:gp_applications_inexpensive}
}
\subfloat[]{%
\includegraphics{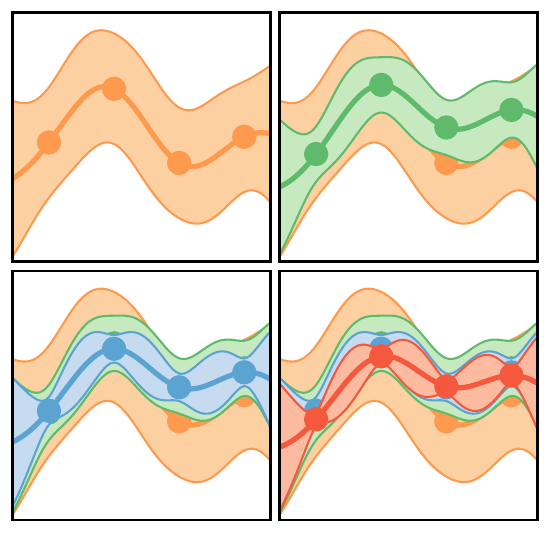}
\label{fig:gp_applications_expensive}
}
\subfloat[]{%
\includegraphics{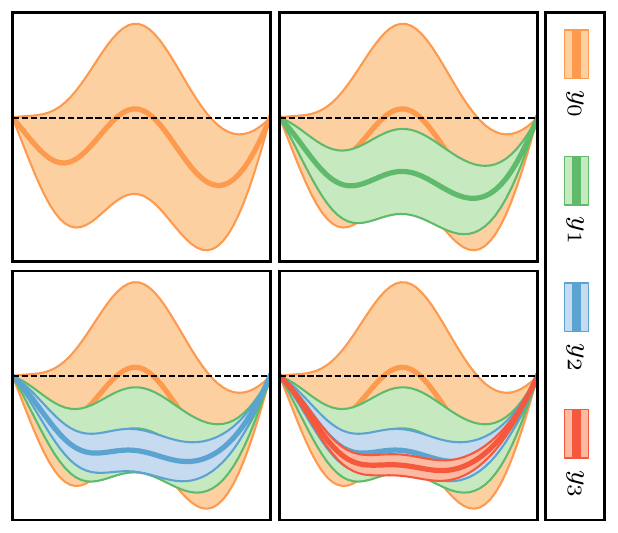}
\label{fig:gp_applications_constraint}
}
\caption{
Toy predictions of $\genobs(\kinparvec) = \genobs_n(\kinparvec) + \delta\genobs_n(\kinparvec)$ vs.\ $\kinparvec$ for $n \leq 3$.
Shaded regions denote $2\sigma$ bands, whose colors correspond to orders described in the legend.
True values for $\param$ and $Q$ are assumed throughout. (a) A computationally inexpensive system with truncation error uncertainty.
Higher-order predictions are generally contained within the lower-order bands.
(b) A computationally expensive system that combines both truncation error and interpolation error quickly and analytically.
Three data points from (a) were used to fit $\genobs_n(\kinparvec)$ at each order, with zero numerical noise assumed.
Solid lines are the interpolants, which are close to the $\genobs_n(\kinparvec)$ from (a) to within their bubble-like interpolation errors.
Note that uncertainty in $\genobs_3(\kinparvec)$ is almost purely due to interpolation. (c) An inexpensive system with the corrections $c_n(\kinparvec)$ constrained to be zero at the endpoints.
This imposes a constraint on $\delta\genobs_n(\kinparvec)$ as well.
As expected, the truncation error smoothly vanishes when approaching the endpoints.}
\label{fig:gp_applications}
\end{figure*}

\textbf{Inexpensive Predictions:} If predictions are inexpensive, then $\genobs_k(\kinparvec)$ can be computed at all $\kinparvec$ of interest.
In this case, we do not need to interpolate $\genobs_k(\kinparvec)$ and can instead focus on estimating the truncation error, i.e., estimating $\param$ and $Q$.
In this data-rich environment, we must be careful when learning the point estimates for these quantities, since too much data within one correlation length $\ell$ can cause ill-conditioned matrices and thus poor point estimates~\cite{Ababouconditionnumbercovariance1994,AndrianakiseffectnuggetGaussian2012,Mohammadianalyticcomparisonregularization2016,Pepelyshevrolenuggetterm2010,RanjanComputationallyStableApproach2011}.
We suggest simply choosing a representative subset of data for learning $\param$ and $Q$ and using a small nugget to regularize the matrix inversion.
Then $\delta\genobs_k$ can be produced at \emph{all} points of interest without issue using Eq.~\eqref{eq:discr_k_prior}.
The distribution for the full observable $\genobs = \genobs_k + \delta\genobs_k$ has mean and covariance function given by
\begin{align}
    \muth(\kinparvec) & = \genobs_k(\kinparvec) + \discrmean{k}(\kinparvec) \label{eq:muth_inexpensive} \\
    \covth(\kinparvec, \kinparvec'; \ell) & = \sdth^2 \discrcorr{k}(\kinparvec, \kinparvec'; \ell). \label{eq:covth_inexpensive}
\end{align}
An example of the posteriors for $\genobs(\kinparvec)$ assuming different max orders $k$ is given in Fig.~\subref*{fig:gp_applications_inexpensive}.

\textbf{Expensive Predictions:} If predictions are expensive, then one may want an estimate of $\genobs_k(\kinparvec)$ at some $\kinparvec$ that was not explicitly computed with the EFT\@.
Thus, we now have both interpolation error and truncation error to worry about.
Fortunately, GPs are \emph{made} for interpolation and are very fast.
In this case, we need a distribution for $\genobs_k(\kinparvec)$ as opposed to $\delta\genobs_k(\kinparvec)$.
Following Eqs.~\eqref{eq:Xk}, \eqref{eq:cn_iid}, and~\eqref{eq:normal_sum}, one can write
\begin{align} \label{eq:genobs_k_prior}
    \genobs_k(\kinparvec) \given \param, Q \sim \GP[\genobsmean{k}(\kinparvec), \sdth^2\genobscorr{k}(\kinparvec, \kinparvec'; \ell)],
\end{align}
where
\begin{align}
    \genobsmean{k}(\kinparvec) & \equiv  \genobsref(\kinparvec)\frac{1 - Q(\kinparvec)^{k+1}}{1 - Q(\kinparvec)}\mu \equiv \genobsbasisfunctrans{k}(\kinparvec)\mu  \label{eq:genobsk_basisfunc} \\
    \genobscorr{k}(\kinparvec, \kinparvec';\ell) & \equiv \genobsref(\kinparvec)\genobsref(\kinparvec') r(\kinparvec, \kinparvec';\ell) \notag \\
    & \times \frac{1 - [Q(\kinparvec)Q(\kinparvec')]^{k+1}}{1 - Q(\kinparvec)Q(\kinparvec')}. \label{eq:genobsk_corrfunc}
\end{align}
Now it is a simple matter to estimate $\param$ and $Q$ as before, and then apply Eqs.~\eqref{eq:gp_conditional_pred}--\eqref{eq:interp_cov} to Eq.~\eqref{eq:genobs_k_prior}.
Because the conditional distribution of Eq.~\eqref{eq:gp_conditional_pred} is \emph{still Gaussian}, one can again use the fact that GPs are closed under addition to compute $\genobs = \genobs_k + \delta\genobs_k$, which has mean and covariance functions
\begin{align}
    \muth(\kinparvec) & = \tildegenobsmean{k}(\kinparvec) + \discrmean{k}(\kinparvec) \label{eq:muth_expensive} \\
    \covth(\kinparvec, \kinparvec'; \ell) & = \sdth^2 \left[\tildegenobscorr{k}(\kinparvec, \kinparvec'; \ell) +  \discrcorr{k}(\kinparvec, \kinparvec'; \ell) \right]. \label{eq:covth_expensive}
\end{align}
[Remember that the tildes above $m$ and $R$ refer to the conditioning shown in Eqs.~\eqref{eq:gp_conditional_pred}--\eqref{eq:interp_cov}.]
See Fig.~\subref*{fig:gp_applications_expensive} for an example when only a few points are computed at each order and used to estimate the full prediction with combined interpolation errors and truncation errors.

\textbf{Symmetry Constraints:}
Above we discussed how to constrain $\genobs_k(\kinparvec)$ based on known data, but sometimes we could constrain $\delta\genobs_k(\kinparvec)$ based on, e.g., symmetry arguments.
That is, we may know that higher-order coefficients are zero at a particular $\kinparvec$.
In this case, we need only constrain $\delta\genobs_k(\kinparvec)$ via Eqs.~\eqref{eq:gp_conditional_pred}--\eqref{eq:interp_cov} according to our knowledge of the system.
This could apply regardless of whether $\genobs_k(\kinparvec)$ is expensive or inexpensive.
Again, since Eq.~\eqref{eq:gp_conditional_pred} is still Gaussian, $\genobs = \genobs_k + \delta\genobs_k$ is a GP.
One need only make the replacements $\discrmean{k} \to \tildediscrmean{k}$ and $\discrcorr{k} \to \tildediscrcorr{k}$ in Eqs.~\eqref{eq:muth_inexpensive}--\eqref{eq:covth_inexpensive} and~\eqref{eq:muth_expensive}--\eqref{eq:covth_expensive}.
Figure~\subref*{fig:gp_applications_constraint} shows an example of how information about symmetries can be applied to an inexpensive system.

\textbf{EFT Fitting:}
Our model permits---for the first time---a rigorous inclusion of all higher-order terms and their correlations within the fitting procedure.
Interpolation of expensive predictions fits seamlessly into this framework, which allows more experimental data to be used than might otherwise be possible.
The posterior for the LECs $\lecs_k$ of the $k$th order EFT, given data for different observables $\data = \{\genobsexpset\}_i$, is given by Bayes' theorem
\begin{align}
    \pr(\lecs_k \given \data, I) & \propto \pr(\data \given \lecs_k, I) \pr(\lecs_k) \notag \\
    & = \pr(\lecs_k) \prod_i \pr(\data_i \given \lecs_k, I_i)
\end{align}
where $I_i$ is the prior information for the $i$th observable, and we have assumed that different observables are conditionally independent of one another given $\lecs_k$.

Consider one specific observable $\genobs$.
Prior information about $\genobs$ could include $\param$ and $Q$ (from previous lower-order fits of the EFT, for example) and also its experimental uncertainty $\covexp$.
The values of $\lecs_k$, $\param$, and $Q$ are sufficient to construct the distribution for $\genobs(\kinparvec;\lecs_k) = \genobs_k(\kinparvec;\lecs_k) + \delta\genobs_k(\kinparvec)$, which is a GP with mean function $\muth(\kinparvec;\lecs_k,\param,Q)$ and covariance function $\covth(\kinparvec,\kinparvec';\lecs_k,\param,Q)$.
The exact form of these functions is dependent on whether the predictions are interpolated and whether the truncation error is constrained, as discussed above.
Once the distribution for $\genobs(\kinparvec;\lecs_k)$ is determined, then the likelihood for this observable is given by
\begin{align} \label{eq:one_obs_parameter_estimation}
    \pr(\data_i \given \lecs_k, I_i) \propto e^{-\frac{1}{2}\transpose{(\genobsexpset - \muthset)} (\covth + \covexp)^{-1} (\genobsexpset - \muthset)}.
\end{align}
This is an extension of the likelihood proposed in Ref.~\cite{WesolowskiExploringBayesianparameter2019}, whose ``uncorrelated'' and ``fully correlated'' models correspond to $\ell=0$ and $\ell=\infty$ limits.
Applying Eq.~\eqref{eq:one_obs_parameter_estimation} to data with finite positive $\ell$ is a subject for future work.

%%%%%%%%%%%%%%%%%%%%%%%%%%%%%%%%%%%%%%%%%

\section{Model checking}
\label{sec:model_checking}

Though the application of our model to any given dataset is simple enough in practice, one must validate that the modeling assumptions made in Sec.~\ref{sec:the_model} are appropriate for the system under consideration.
If this model is inappropriate, then the estimates of $\param$ and $Q$, along with the truncation error distributions, could be biased or even meaningless.
Conversely, if one assumes our model describes how an EFT \emph{should} behave, then it would be of great interest to know when a given EFT is failing.
Thus a well chosen set of model checking diagnostics are critical to test our assumptions against the data at hand.

To test our assumptions against reality, we must first enumerate them:
\begin{enumerate}
  \item The coefficients $c_n$ are \iid\ realizations of a GP.
  \item The $c_n$ GP's mean and covariance functions---and their parameters $\param$---have been correctly identified.
  \item From the knowledge of a few coefficients, we can create a statistically meaningful distribution for the truncation error.
\end{enumerate}
Model-checking diagnostics can assess whether these assumptions are violated at a statistically significant level.
The question of how to diagnose issues with GP fitting and prediction has been addressed thoroughly in Ref.~\cite{BastosDiagnosticsGaussianProcess2009}.
However, our situation is novel compared to that of the everyday GP practitioner: (1) they typically have points along one curve, we have multiple \iid\ curves; (2) their main focus is often the accuracy of the GP regressor's predictions, our focus is mainly on the parameters $\param$ themselves; and (3) our model for the $c_n$ leads to a truncation error distribution which itself can be tested.
Nevertheless, the same techniques discussed in 
Ref.~\cite{BastosDiagnosticsGaussianProcess2009} are directly applicable here.

Throughout the discussion below, we assume that the parameters for a GP have either been fixed \emph{a priori}, or calibrated/marginalized via training data $\ftrain$.
The resulting GP (or TP) is then to be evaluated against test (or validation) data $\fvalid$.
One should take care that the training set $\ftrain$ does not overlap with any points in $\fvalid$.
We have greater than average freedom with how we partition $\ftrain$ and $\fvalid$ because our data contain multiple \iid\ curves $c_n$ that could be tested against one another and because our truncation error model can be tested against reality.
For now, assume that $\ftrain$ contains data on \emph{one} curve and was used to construct an \emph{emulator} via Eqs.~\eqref{eq:gp_conditional_pred}--\eqref{eq:interp_cov} to predict $\fvalid$ from that same curve.
We will return to the details on how one might choose different emulators, or how to partition the training/test sets differently, in Sec.~\ref{sub:honorable_mentions}.
Below, we use the generic notation of $\inputvec{m}$ and $K$ for the mean and covariance of the emulator (with $\hypdf$ degrees of freedom in the TP case).

\subsection{Mahalanobis Distance}
\label{sub:mahalanobis_distance}

The common way to measure the loss, or incorrectness, of a prediction is through the sum of squared residuals.
This metric as a summary assumes that the errors at different input points $\kinparvec$ are completely uncorrelated; the individual weighted residuals at each point are simply summed up.
The Mahalanobis distance is the multivariate analog of this idea that takes into account these correlations.
That is,
\begin{align}
    \DMD^2(\fvalid) = \transpose{(\fvalid - \inputvec{m})} K^{-1} (\fvalid - \inputvec{m}).
\end{align}
Large $\DMD^2$ implies that the validation data does not match the emulator.
But what defines ``large'' and ``small'' in this context?
A necessary component of any diagnostic is a \emph{reference distribution}.
A reference distribution sets the scale of variation and defines what normal and abnormal diagnostic values look like.

Suppose that we have validation data at $M$ locations.
Then, if the emulator is Gaussian, the reference distribution for $\pr(\DMD^2)$ is a $\chi^2$ distribution with $M$ degrees of freedom.
If the emulator is a Student-$t$ distribution with $\nu$ degrees of freedom, the reference distribution is an $F$ distribution $F_{M,\nu}$ that is scaled by $s = (\nu-2)M/\nu$~\cite{BastosDiagnosticsGaussianProcess2009}.
(That is, $\DMD^2 / s$ follows a standard $F_{M,\nu}$ distribution.)
The reference distributions in either the GP or TP cases do not depend on the specific form of the means and covariance functions: only the number of validation data and, if applicable, the degrees of freedom.
Both the $\chi^2$ and the $F$-distributions are available in standard statistical libraries, such as SciPy, and permit simple analytic evaluations of credible intervals.

But, in general, the distributions of such diagnostics may not have a simple form.
A surefire way to generate a distribution in any case is to simply sample from the emulator and compute the diagnostic for each $\funcvec$ in the sample.
Then the validation data $\fvalid$ could then be rejected at a 68\% or 95\% level, for example, based on the empirical reference distribution.

\subsection{Pivoted Cholesky Decomposition}
\label{sub:variance_decompositions}

The MD is great as a one-number summary, but more information can be gained by considering decompositions of this quantity.
Let $G$ be defined by $K = G \transpose{G}$.
Then
\begin{align}
    \DVARIT{G} = G^{-1} (\fvalid - \inputvec{m})
\end{align}
is a variance decomposition of $\DMD^2 = \DVARIT{G}^\trans \DVARIT{G}$.
There is freedom in the exact definition of this diagnostic because $G$ is not unique.
The \emph{pivoted} Cholesky decomposition of $K$ leads to a diagnostic ($\DVAR{PC}$) that pinpoints the data contributing to a failing $\DMD^2$ and shows useful patterns when plotted graphically vs.\ index~\cite{BastosDiagnosticsGaussianProcess2009}.
A misestimated variance leads to unusually sized $\DVAR{PC}$ across all indices, whereas an incorrect estimation of the length scale leads to failing $\DVAR{PC}$ at large index.
% For example, when plotted vs.\ index, large or small values at small index indicate non-stationarity or a misestimation of the predictive variance.
% Additionally, abnormally small or large values at large index indicate that the correlation parameters are misestimated or that the chosen covariance function is incorrect.

The reference for each component of $\DVARIT{G}$ is distributed as a standard Gaussian (for a GP emulator) since the points have been scaled and de-correlated by $G$.
For the TP emulator, the reference is a Student-$t$ distribution with $\hypdf$ degrees of freedom.
The desired credible interval (here, $2\sigma$) can be computed accordingly.

\subsection{Credible Interval Diagnostic}
\label{sub:ci_diagnostic}

The credible interval diagnostic tests the accuracy of the emulator.
An emulator is accurate if a $100\alpha\%$ credible interval (CI) contains approximately $100\alpha\%$ of the validation data.
This diagnostic could be particularly useful when comparing experimental data $\genobsexpset$ to the full prediction with truncation uncertainty, since it captures whether the error behaves as advertised.
Following Ref.~\cite{BastosDiagnosticsGaussianProcess2009}, start by building $\CI_i(\alpha)$, the $100\alpha\%$ marginal CI at the point $\kinparvec_i$, for each of the $M$ corresponding $\func_i \in \fvalid$.
Then compute the accuracy of the credible intervals:
\begin{align}
  \DCI(\alpha, \fvalid) = \frac{1}{M}\sum_{i=1}^M \indicator[\func_i \in \CI_i(\alpha)],
\end{align}
where $\indicator$ is an indicator function, i.e., it equates to 1 if its argument is true and equates to 0 otherwise. Note that although the $\CI$ may be informed by training data at many points, the $\DCI$ is computed point by point.

This diagnostic is the analog to the ``consistency plots'' shown in Ref.~\cite{MelendezBayesiantruncationerrors2017}, which used the pointwise truncation model.
It was shown that in this pointwise case, where the data are assumed uncorrelated with one another, the reference distribution is a simple binomial.
Since the data here are curves, the reference is more complicated.
Here we compute the reference distribution by simulation, that is, we sample the emulator and compute $\DCI$ for each sample.
The creation of a reference in this manner takes into account correlations by widening the acceptable region of accuracy, though the diagnostic is still inherently a point-by-point summary.
Interestingly, we have found that the simulated reference is approximately a binomial with an effective number of data $M_{\text{eff}} = \Delta \kinparvec / \ell$, where $\Delta\kinparvec$ is the difference between the minimum and maximum values of $\kinparvec$.

\subsection{Honorable Mentions}
\label{sub:honorable_mentions}

We have tested many more diagnostics than are shown graphically in this work, such as
\begin{itemize}
    \item Variogram: A common tool used to investigate correlation structure by looking at the variability of pairs of points across the domain~\cite{CressieFittingvariogrammodels1985,BowmanInferencevariograms2013}.
    We have found that it is too variable for our application.
    \item KL Divergence: Another one-number summary for emulators. We did not find that it gave any new information beyond $\DMD^2$.
    \item Other variance decompositions: In addition to the $\DVAR{PC}$, we considered the standard Cholesky decomposition and the eigendecomposition. The Cholesky decomposition has the benefit that each index corresponds to a single validation datum, but it does not order the indices in a visually intuitive manner for diagnosing problems. The eigendecomposition, on the other hand, displays similar qualities as the pivoted Cholesky decomposition when plotted vs.\ index, but its indices do not correspond to individual validation points. Thus, we choose to display only the pivoted Cholesky here, which enjoys the best of both worlds.
\end{itemize}
The above diagnostics are still available in the \texttt{gsum}~\cite{gsum} package despite not appearing here.

\subsection{How to Choose an Emulator}
\label{sub:emulator_choices}

We have described the most common approach for diagnosing GP emulators: $\ftrain$ and $\fvalid$ are function values from one curve evaluated at mutually exclusive sets of input points $\kinparvec$.
Now we discuss alternatives to this approach based on the questions our diagnostics are designed to address.
This discussion is split into two parts: (1) evaluating coefficients $c_n$ and (2) evaluating predictions $\genobs_n$.

By testing the coefficients we can address the question of whether the observed convergence pattern matches our assumptions.
Because they are assumed to be \iid, the parameters of their process can be learned from training data along each $c_n$ simultaneously or along a subset of the curves.
Then an emulator for some $c_n$ could be used to predict test data between the interpolating points, regardless of whether those points along $c_n$ were used to estimate $\param$.

A problem occurs when using interpolating processes if the training data are too close~\cite{Ababouconditionnumbercovariance1994,AndrianakiseffectnuggetGaussian2012,Mohammadianalyticcomparisonregularization2016,Pepelyshevrolenuggetterm2010,RanjanComputationallyStableApproach2011}.
The bubble-like interpolation uncertainty becomes very small, and can approach the size of the nugget used for numerical stability.
In this case the diagnostics become highly sensitive to the nugget regulator, a clear sign of diagnostics gone awry.
One way to avoid this issue is to spread the training points out relative to the length scale of the system, but this solution comes with a price.
The estimates of $\param$ and the accuracy of the interpolants will suffer.

But the issue of numerical instability can be mitigated by avoiding interpolating processes altogether.
Our goal is to test whether the $c_n$ are drawn from one underlying process, so why not test the coefficients against this process rather than interpolating?
The marginal variance of the underlying process should always be large with respect to the nugget size, which sidesteps the problem of comparable sizes.
Because the emulator is not forced to interpolate the training data (only $\param$ is tuned), it may be less crucial to exclude training data from the validation set and would allow for more data to be used in the analysis.
Diagnosing against the underlying process is thus the method of choice in this work.

Now we outline the diagnosis procedure for  $\genobs_n$.
The predictions $\genobs_n$ allow us to test whether our truncation (and possibly interpolation) model behaves as advertised.
The data used to estimate the parameters $\param$ could include all EFT predictions up to the highest order available, with the caveat that data that are too closely spaced could cause numerical instability in their estimates.
Then an emulator for the full prediction can be constructed as discussed in Sec.~\ref{sub:application_types}, and the validation data could then consist of experimental data $\genobsexpset$.
If testing the truncation error, there is no issue computing diagnostics with the same $\kinparvec$ as was used in the training set.

Modulo the possible numerical issues discussed above, poorly behaving diagnostics are a sign that something is wrong with the EFT or the convergence model assumptions.
To diagnose such problems requires a working knowledge of what success and failure look like.
Hence, it will be more illuminating to see the diagnostics in action rather than to further discuss theory.
After the following illustrations it should be clear that statistics is a powerful tool to diagnose physics and EFT failure modes.

\section{Toy Application}
\label{sec:toy_application}

Figure~\ref{fig:preds_to_coeffs} is a summary of how one takes the theoretical predictions of an EFT and extracts coefficients, which then inform the truncation error model as described in Sec.~\ref{sec:the_model}.
This section takes the predictions from Fig.~\subref*{fig:pred_to_coeffs_preds} as given, and provides an example workflow for extracting physical insights from the data.
% Toy data helps make the results interpretable and allows comparisons to the generating process.
By working with toy predictions, we are able to assess the accuracy of both the truncation error bands and the extracted GP parameters $\param$.
We test the distribution of the coefficients in Fig.~\subref*{fig:pred_to_coeffs_coeffs} against our GP model, and then compute the truncation error distributions for each $\genobs_n$ in Fig.~\subref*{fig:pred_to_coeffs_preds}, followed by an assessment against a to-all-orders EFT prediction.
Throughout, we show visually how the diagnostics from Sec.~\ref{sec:model_checking} behave and how to know when they are failing. 
% Algorithm~\ref{algo_truncation} is a condensed version of this section.

Let us assume that mapping from predictions $\genobs_n$ to coefficients $c_n$ is well understood, that is, the values of $\genobsref$ and $Q$ are known.
(We will return later to relax this assumption.)
Then we are left with two projects: estimating the process from whence the coefficients were drawn, and subsequently using that process to compute the truncation error distribution, all the while validating assumptions along the way.
The first step is to outline the priors on the GP parameters $\param$.
Here we take $\hypprior\hypm=0$ and $\hypprior\hypdisp=0$ (i.e.\ the mean of the coefficients is known to be zero), which corresponds to the belief that the coefficients are just as likely to be positive as negative and matches the distribution from which they are drawn.
The prior on the variance $\sdth^2$ is given $\hypprior\hypdf = 1$ and $\hypprior\hyptau^2 = 1$, which is fairly uninformative with a mode at $1/3$ and a very heavy tail.
Finally, an improper uniform prior on the interval $(0, \infty)$ is taken for the length scale $\ell$.

\begin{figure*}
\captionsetup[subfloat]{margin=.5pt,captionskip=-14pt}
\subfloat[]{%
\includegraphics{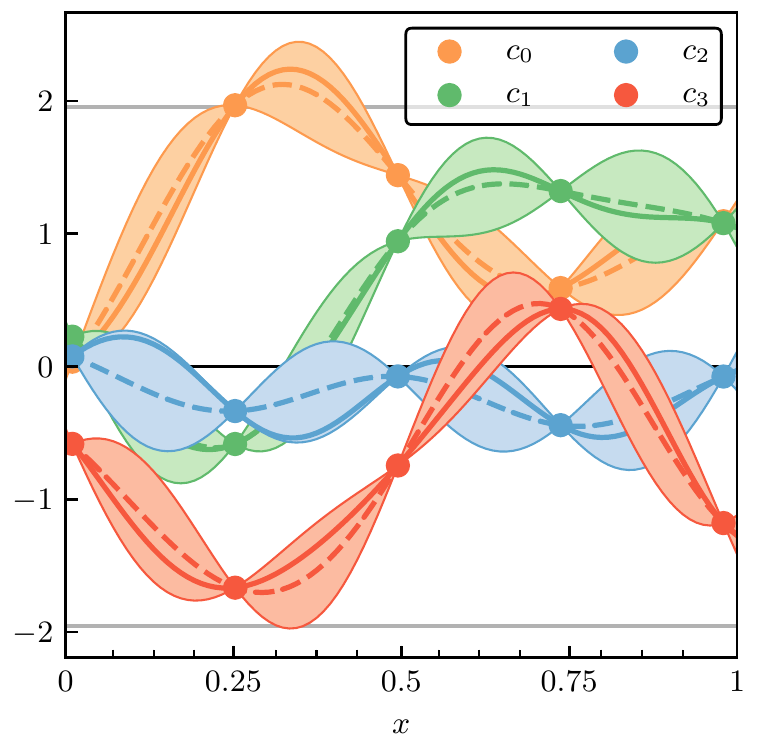}
\label{fig:toy_coeffs_and_diagnostics_coeffs}
}
\subfloat[]{%
\includegraphics{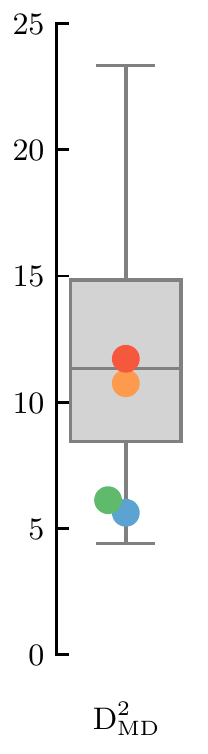}
\label{fig:toy_coeffs_and_diagnostics_md}
}
\subfloat[]{%
\includegraphics{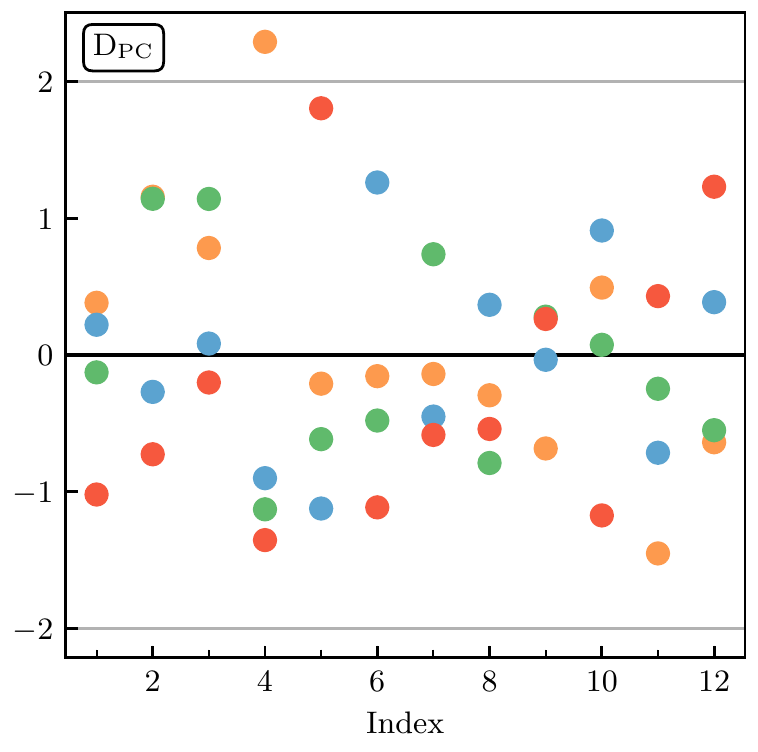}
\label{fig:toy_coeffs_and_diagnostics_pc}
}
\caption{Toy coefficients and their corresponding diagnostics computed against the underlying process. Colors match across each subplot. (a) The coefficients from Fig.~\ref{fig:preds_to_coeffs} and their corresponding GP emulators. The mean and $2\sigma$ bands of the calibrated underlying process are shown as black and gray lines, respectively. The colored dashed lines and corresponding bands are the interpolants and $2\sigma$ bands found by fitting this underlying GP to each $c_n$ individually with Eqs.~\eqref{eq:interp_mean}--\eqref{eq:interp_cov}.
The training data are denoted by dots, whereas the validation data locations are indicated by the minor ticks.
(b) The Mahalanobis distance computed against the \emph{underlying} (not interpolating) process.
The interior line, box end caps, and whiskers on the box plot show the median, 50\% credible intervals, and 95\% credible intervals, respectively.
The blue $\DMD^2$ for $c_2$ is the smallest, which shows that the diagnostic reflects our intuition as discussed in Fig.~\ref{fig:preds_to_coeffs}.
Overall, the $\DMD^2$ are reasonably sized.
(c) The pivoted Cholesky diagnostic $\DVAR{PC}$ vs.\ index, with gray lines that represent its $2\sigma$ error bands.
The points seem to be distributed as expected: approximately distributed as a standard Gaussian, with no real pattern vs.\ index.
}
\label{fig:toy_coeffs_and_diagnostics}
\end{figure*}

\begin{figure*}
\subfloat[]{%
\includegraphics{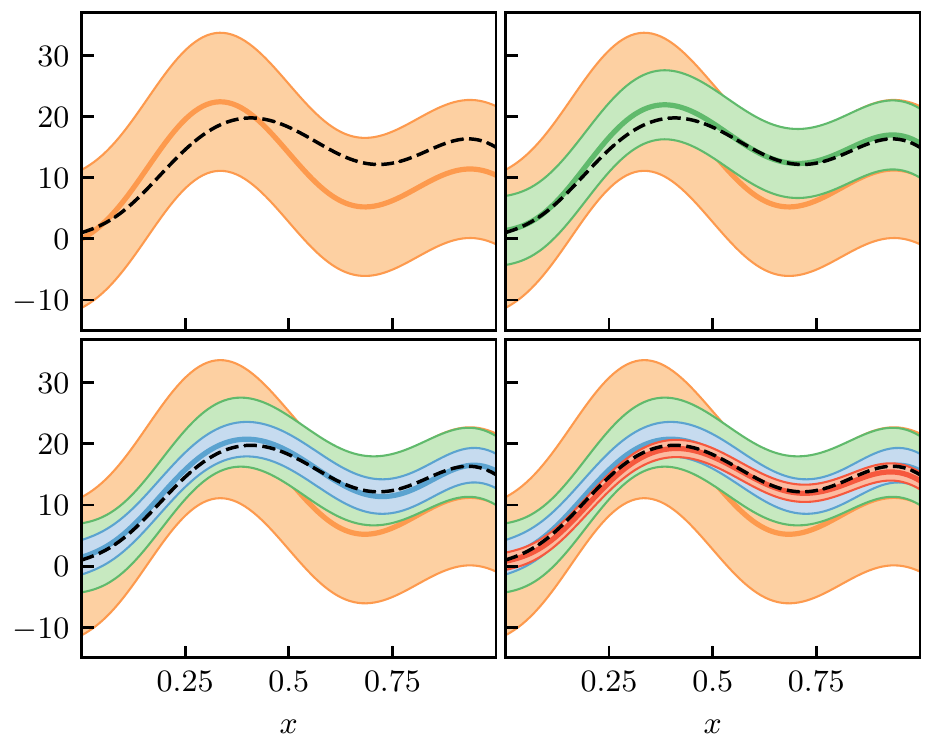}
\label{fig:toy_truncation_estimates_bands}
}
\subfloat[]{%
\includegraphics{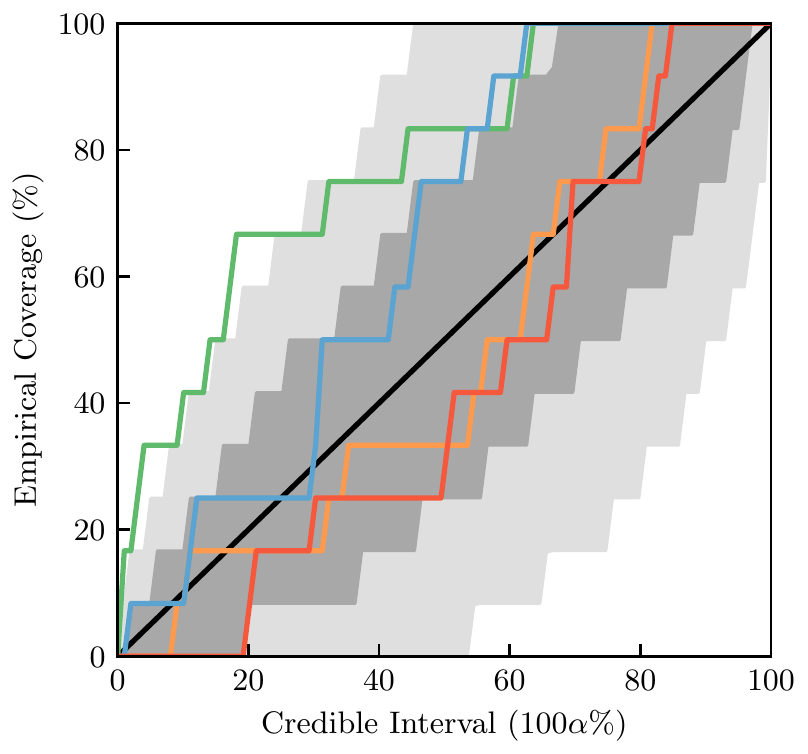}
\label{fig:toy_truncation_estimates_diagnostics}
}
\caption{(a) Toy predictions from Fig.~\ref{fig:preds_to_coeffs} and their corresponding $2\sigma$ truncation error estimates generated from the underlying process in Fig.~\ref{fig:toy_coeffs_and_diagnostics}. The black dashed line is the observable computed up to order $n=20$, which represents the true value of the observable and is used as validation data.
Minor ticks denote the validation data locations used in the diagnostic.
(b) The credible interval diagnostic for the truncation error bands.
The stepwise nature of the lines arises from the finite number of points used in the diagnostic.
Dark and light bands represent $1\sigma$ and $2\sigma$ credible intervals.
By chance, $\genobs_1$ is very close to the true curve, which results in a $\DCI$ that is too accurate at small $\alpha$.
}
\label{fig:toy_truncation_estimates}
\end{figure*}

\begin{table}[tb]
\caption{Parameter estimates for the emulators in Fig.~\protect\subref*{fig:toy_coeffs_and_diagnostics_coeffs} using a nugget of $\sdnugget^2 = 10^{-10}$.
}
\label{tab:toy_data_parameters}
% \renewcommand{\arraystretch}{1.4}%
% \setlength\extrarowheight{1pt}
% Use S prefix to enable the cellspace package
% which makes the spacing look nice.
\begin{ruledtabular}
\begin{tabular}{ScScScSc}
  $\param$ & Prior & MAP & True \\
  \colrule
  $\mu$ & $\normal(0,0)$ & 0 & 0 \\
  $\sdth$ & $\invchisq(1,1)$ & 0.97 & 1 \\
  $\ell$ & $U(0,\infty)$ & 0.20 & 0.2 \\
  Q & $U(0,1)$ & 0.48 & 0.5 \\
\end{tabular}
\end{ruledtabular}
\end{table}

\begin{figure}
\includegraphics{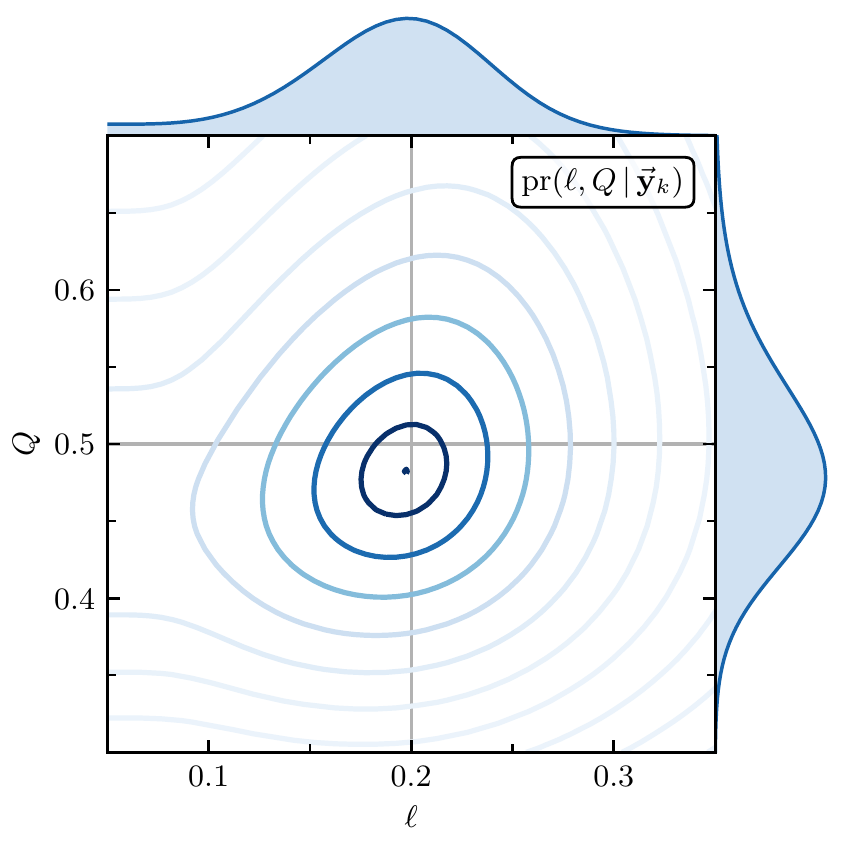}
\caption{The joint and marginal posteriors for $\ell$ and $Q$ given data in Fig.~\protect\subref*{fig:toy_coeffs_and_diagnostics_coeffs}, with a uniform prior $\pr(\ell, Q) \propto 1$.
The contours increment in approximately half-standard-deviation intervals, with a point denoting the MAP value. The gray lines represent the true values of $\ell$ and $Q$.
The training data are spaced with $\Delta \kinparvec \approx 0.25$, meaning that the $\ell$ posterior cannot entirely discount the possibility of small $\ell$ without prior information.
}
\label{fig:toy_Q_ell_posterior}
\end{figure}

\begin{figure*}[tb]
\captionsetup[subfloat]{margin=.5pt,captionskip=-14pt}
\subfloat[]{%
\includegraphics{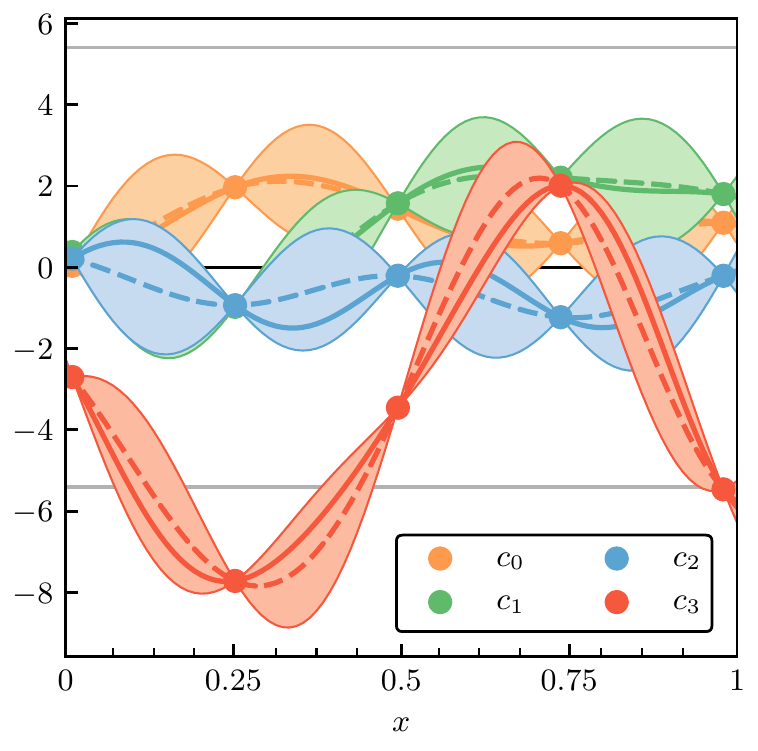}
\label{fig:toy_coeffs_and_diagnostics_coeffs_fail}
}
\subfloat[]{%
\includegraphics{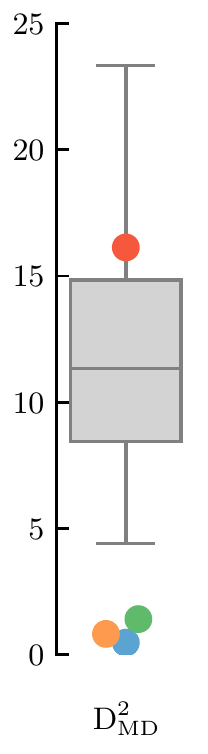}
\label{fig:toy_coeffs_and_diagnostics_md_fail}
}
\subfloat[]{%
\includegraphics{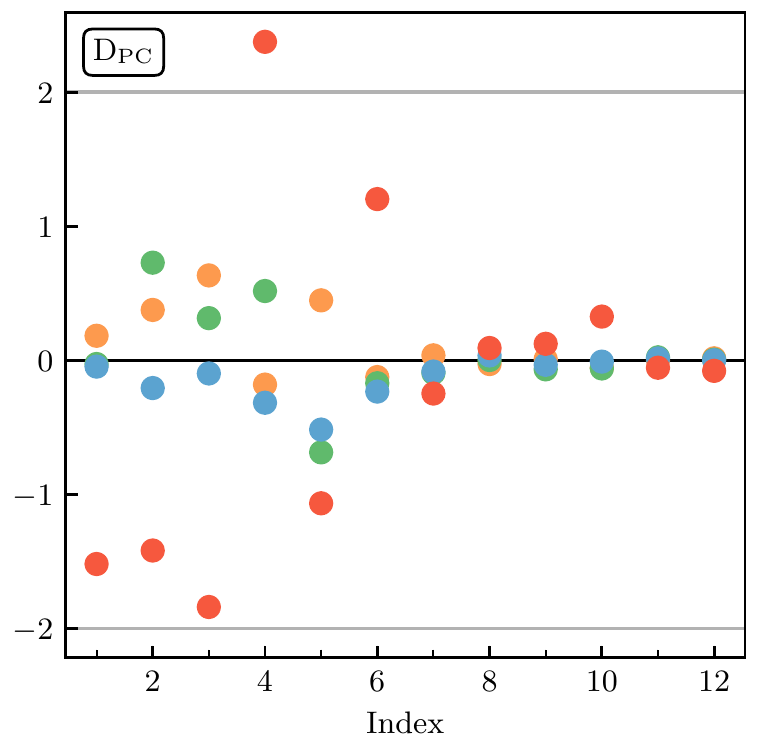}
\label{fig:toy_coeffs_and_diagnostics_pc_fail}
}
\caption{Toy coefficients extracted using $Q=0.3$ (opposed to the true $Q=0.5$) and their corresponding diagnostics.
In (a), the training data are denoted by dots, whereas the validation data locations are indicated by the minor ticks.
See Fig.~\ref{fig:toy_coeffs_and_diagnostics} for the remaining figure notation.
The coefficients no longer look particularly \iid\ given the large size of $c_3$.
This biases the estimates of $\sdth^2$ and $\ell$ and results in the failure modes of the Mahalanobis distance and its pivoted Cholesky distribution.
The funnel-like behavior of the pivoted Cholesky when moving from low to high index signals a misspecification of the length scale, which indeed is biased downwards due to the incorrectly chosen $Q$.
}
\label{fig:toy_coeffs_and_diagnostics_small_Q}
\end{figure*}

Once priors have been assigned to $\param$, the next step is to update our beliefs using the coefficients as data.
We choose 5 points along each of the 4 $c_n$ with $\Delta \kinparvec \approx 0.25$, shown as dots in Fig.~\subref*{fig:toy_coeffs_and_diagnostics_coeffs}, and use them to compute MAP values for $\sdth^2$ and $\ell$ [Eqs.~\eqref{eq:sdth_point_estimate} and~\eqref{eq:ell_given_Q_posterior_corr}].
This optimization procedure results in estimates that are remarkably close to the true parameter values, see Table~\ref{tab:toy_data_parameters}.
Plugging these estimates into Eq.~\eqref{eq:cn_iid}, yields an estimate of the \emph{underlying} process that generated the $c_n$, whose mean and marginal variance are plotted on Fig.~\subref*{fig:toy_coeffs_and_diagnostics_coeffs} as black solid and dashed lines.
The interpolating processes can then be computed for each $c_n$ [Eqs.~\eqref{eq:interp_mean}--\eqref{eq:interp_cov}], whose means and marginal variances are given by colored dashed lines and filled areas in Fig.~\subref*{fig:toy_coeffs_and_diagnostics_coeffs}.

We know that the estimated process is quite close to the generating process, but we have no such guarantee when considering real EFT predictions.
This is where the diagnostics proposed in Sec.~\ref{sec:model_checking} come into play.
We choose to compare the $c_n$ validation data, whose locations are indicated by minor ticks in Fig.~\subref*{fig:toy_coeffs_and_diagnostics_coeffs}, to the underlying GP, as opposed to each interpolating process.
Thus, our diagnostics will indicate whether these data points could have feasibly been drawn from the GP that we have learned from the training points.
The results of the diagnostics are shown in Figs.~\subref*{fig:toy_coeffs_and_diagnostics_md} and~\subref*{fig:toy_coeffs_and_diagnostics_pc}.
As expected, the diagnostics behave quite well because the emulator matches the generating process.
The $\DMD^2$ does return a small value for $c_2$, reflecting our intuition that $c_2$ looks abnormally small compared to the others.
The pivoted Cholesky decomposition vector, plotted vs.\ its index, shows an essentially random variation of points, meaning that the variation of the $c_n$ about the mean are distributed approximately as expected.
Again, $c_2$ is rather small, given that the vertical variation should follow a standard normal, but the others show a mixture of large and small values of $\DVAR{PC}$.
Such diagnostics show that the coefficients from the computed EFT orders conform to the assumptions of our convergence model.

The computed EFT orders are only half of the story; we desire truncation bands that take into account all not yet computed orders.
By inserting point estimates of $\param$ and $Q$ into Eq.~\eqref{eq:discr_k_prior} [or by using its marginalized analog, Eq.~\eqref{eq:discr_k_posterior_student_t}], posteriors for the full EFT summation are just as simple as the posterior for the $c_n$.
The truncation error bands for each $\genobs_n$, using the point estimates in Table~\ref{tab:toy_data_parameters}, are shown in Fig.~\subref*{fig:toy_truncation_estimates_bands}.
As an example, a higher-order prediction $\genobs_{20}$ is taken as the true value of the prediction.
Given the truncation error processes and the validation data of $\genobs_{20}$, we can again compute diagnostics to determine whether the truncation errors behave as expected.
We show the credible interval diagnostic in Fig.~\subref*{fig:toy_truncation_estimates_diagnostics}, which assesses the accuracy of the truncation error credible intervals.
On average, the error estimates contain the true value of $\genobs$ in the proportion dictated by the credible interval.
This is the sign of a healthy EFT convergence pattern.

Thus far we have assumed for simplicity that the pathway from order-by-order predictions to naturally sized coefficients, via Eq.~\eqref{eq:constraints}, was known in advance.
But first there is some work to do to even make this conversion:
assumptions must be made for $\genobsref$ and $Q$.
As we have stated previously, $\genobsref$ can be derived based on dimensional analysis for the observable at hand, and $Q$ can be based on a separation-of-scales argument specific to the EFT in use.
But this can be easier said than done.
We have found that one of the most informative pathways to physics insight is graphically exploring a dataset.
Plotting coefficients generated from different assumptions on $\genobsref$ and $Q$ can show very different patterns, some of which better conform to the assumption that the $c_n$ are identically distributed.
Empirical exploration of coefficients' behavior can often lead to useful insights regarding the parameters 
that enter the EFT convergence model. For example, the assignment $\genobsref = 1$ for spin observables in Ref.~\cite{MelendezBayesiantruncationerrors2017} was first hit upon through such exploration.
This process should be iterative, with scale arguments from the EFT reinforcing or critiquing such empirical observations. The diagnostics we have presented here lend statistical rigor to this process. 

An important tool for exploring the EFT convergence is the model evidence~\cite{WesolowskiParameterEstimationForEFTs2016} $\pr(\genobsvecset \given \ell, Q, \genobsref)$, given by Eq.~\eqref{eq:genobsvec_given_ell_Q} (where dependence on $\genobsref$ is shown explicitly here).
Up to prior factors and an unimportant normalization constant, the model evidence is equivalent to the posterior for $\ell$, $Q$, and $\genobsref$.
Suppose that $\ell$ and $Q$ are both scalar quantities with uniform priors and $\genobsref$ is known.
Then the posterior $\pr(\ell,Q \given \genobsvecset)$ can be computed analytically, as is shown in Fig.~\ref{fig:toy_Q_ell_posterior} and summarized in Table~\ref{tab:toy_data_parameters}.
Figure~\ref{fig:toy_Q_ell_posterior} is a pathway to learning the convergence pattern, and other EFT quantities, through data.
In fact, even if $Q$ is a function of $\kinparvec$ and other parameters, such as the EFT breakdown scale $\Lambda_b$, then the posterior for these parameters is still analytic and can be optimized to find a MAP value [see Eq.~\eqref{eq:breakdown_scale_posterior_corr}].

But choosing $Q$ may not be as simple as finding the best $\Lambda_b$, rather, one may need to decide between a discrete set of functional forms for $Q(\kinparvec)$.
In this case, differing $Q(\kinparvec)$ can be thought of as models to choose between.
Moreover, one could question the choice, or functional form, of $\genobsref$.
The model evidence is useful as a diagnostic tool in this case as well. ($\ell$ could be integrated out numerically, if desired.)
In this case, one can compute Bayes' factors, i.e., ratios of the model evidence for different choices  $(Q, \genobsref)$, to determine the choice statistically favored by the EFT convergence pattern~\cite{trotta_bayes_2008, WesolowskiParameterEstimationForEFTs2016}.

As an example of what can occur when $Q$ is misestimated, we now choose $Q=0.3$ (opposed to the true $Q=0.5$) and repeat the analysis of Fig.~\ref{fig:toy_coeffs_and_diagnostics}.
The results of this analysis are shown in Fig.~\ref{fig:toy_coeffs_and_diagnostics_small_Q}.
Since $Q$ is underestimated, then the $c_n$ grow with $n$, as dictated by Eq.~\eqref{eq:constraints}.
Thus $c_3$ is large compared to the others, and causes the $\sdth$ estimate to increase.
In general, the marginal variance is biased upwards by large outliers.
This is reflected by the Mahalanobis distance only capturing $c_3$, but the other coefficients are abnormally small.
The assumption that the curves are identically distributed has broken down.
The pivoted Cholesky decomposition in Fig.~\subref*{fig:toy_coeffs_and_diagnostics_pc_fail} provides further evidence of this breakdown; the lower-order components are too small compared to the reference distribution.
Moreover, there is a trend of smaller errors near the right side of the chart: evidence that the length scale has been misestimated.
Indeed, the length scale found by optimizing the likelihood is $\ell = 0.17$, small enough compared to the true value of $\ell = 0.2$ to show up in this diagnostic.

%%%%%%%%%%%%%%%%%%%%%%%%%%%%%%%%%%%%%%%%%%%%%%%%%%%%%%%%%%%%%%%

\section{Application to NN scattering with chiral EFT potentials}
\label{sec:NN_scattering}

Now that we have introduced both the truncation error model and its diagnostics, we dedicate this section to a real-world example in low-energy nuclear physics.
In this regime, chiral EFT is a popular tool used to develop a systematically improvable description of nuclear observables.
The most popular version of chiral EFT is the original proposed in Weinberg's seminal works, where a nuclear potential is written down and resummed using the Lippmann-Schwinger (LS) equation~\cite{weinberg_nuclear_1990,weinberg_effective_1991}.
This resummation obscures the power counting that was manifest in the potential.
Nevertheless, for certain regulators, the convergence of nucleon-nucleon scattering observables appears consistent with this power counting~\cite{furnstahl_quantifying_2015,MelendezBayesiantruncationerrors2017}.

But the promising results in Refs.~\cite{furnstahl_quantifying_2015,MelendezBayesiantruncationerrors2017} were obtained using the pointwise convergence model.
Though it permitted a statistical extraction of the EFT breakdown scale $\Lambda_b$, strong conclusions about $\Lambda_b$ could not be made due to the lack of correlations inherent in the pointwise approach.
Here we readdress a subset of these analyses using our GP approach.
A more thorough treatment of this and other systems is a topic for future work.

We consider three neutron-proton ($\npr$) scattering observables: the differential cross section $\sigma(\theta)$ as a function of center-of-mass angle $\theta$, the total cross section $\sigmatot$ as a function of lab energy $\Elab$, and the spin observable $A$ as a function of $\theta$.
Each of the observables are computed using the semilocal potential of Epelbaum, Krebs, and Mei\ss{}ner (EKM)~\cite{epelbaum_precision_2015,epelbaum_improved_2015}. 
We find that, for $NN$ data in chiral EFT, the coefficients look very smooth and center around a mean of zero, see Ref.~\cite{MelendezBayesiantruncationerrors2017}.
Thus, we model the correlation structure using the squared exponential kernel, Eq.~\eqref{eq:se_kernel}, and choose a mean function of zero.

Once the coefficients' mean and covariance functions have been specified, choices must be made for $\genobsref$ and $Q$.
In Ref.~\cite{MelendezBayesiantruncationerrors2017}, $\genobsref = \genobs_0$ was used for the total and differential cross section, while it was argued that $\genobsref = 1$ is the appropriate scale for spin observables.
Here we argue for a slight modification: we choose $\genobsref = \genobs_5$ for the differential cross section since the leading-order prediction $\genobs_0$ gets dangerously close to 0 in some regions, which causes unnatural peaks in the coefficient curves.

The expansion parameter $Q$ is a ratio of low- to high-energy scales.
The high-energy scale is the breakdown scale of the EFT $\Lambda_b$, whose value is obscured by the LS equation.
For $\npr$ scattering in chiral EFT, the low-energy scales include quantities such as the beam kinetic energy in the lab frame $\Elab$ and the pion mass $m_\pi$.
In the center of momentum frame, one can rewrite the lab energy as the relative momentum given the proton mass $M_p$ and neutron mass $M_n$
\begin{align} \label{eq:prel}
  \prel^2 = \frac{\Elab M_p^2(\Elab + 2M_n)}{(M_p+M_n)^2 + 2M_p\Elab}.
\end{align}
With $\prel$ a function of $\Elab$, the expansion parameter is defined as
\begin{align}
  Q(\Elab) = \frac{f(\prel,m_\pi)}{\Lambda_b}
\end{align}
where $f(\prel,m_\pi)$ is some mapping of $\prel,m_\pi$.

Again, due to the resummation performed by the LS equation, the exact mapping $f(\prel,m_\pi)$ is not known.
One would expect that $\lim_{\prel \gg m_\pi} f(\prel,m_\pi)=\prel$ and $\lim_{\prel \ll m_\pi} f(\prel,m_\pi) = m_\pi$, but many functional forms could satisfy these constraints.
In prior work, we have assumed the mapping $f(\prel,m_\pi) \to \max(\prel,m_\pi)$, or a smoothed max function
\begin{align} \label{eq:Q_smooth_max}
  f(\prel,m_\pi) \to \frac{\prel^n + m_\pi^n}{\prel^{n-1}+m_\pi^{n-1}}
\end{align}
with $n=8$~\cite{MelendezBayesiantruncationerrors2017}.
Here we choose the smooth max function, Eq.~\eqref{eq:Q_smooth_max}.

\begin{figure*}
\captionsetup[subfloat]{margin=0pt}
\subfloat[]{%
\includegraphics{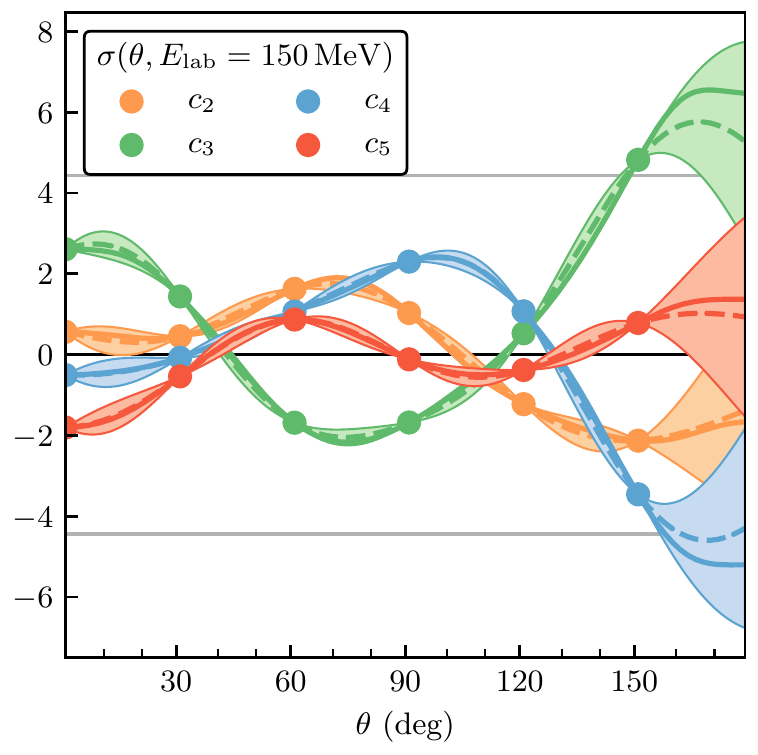}
}
\subfloat[]{%
\includegraphics{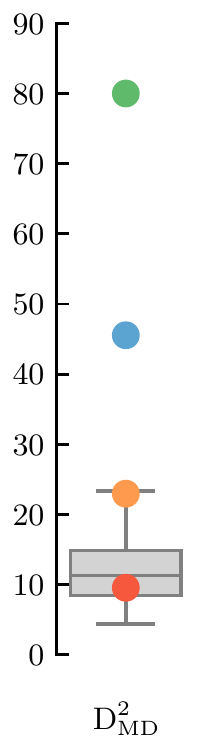}
}
\subfloat[]{%
\includegraphics{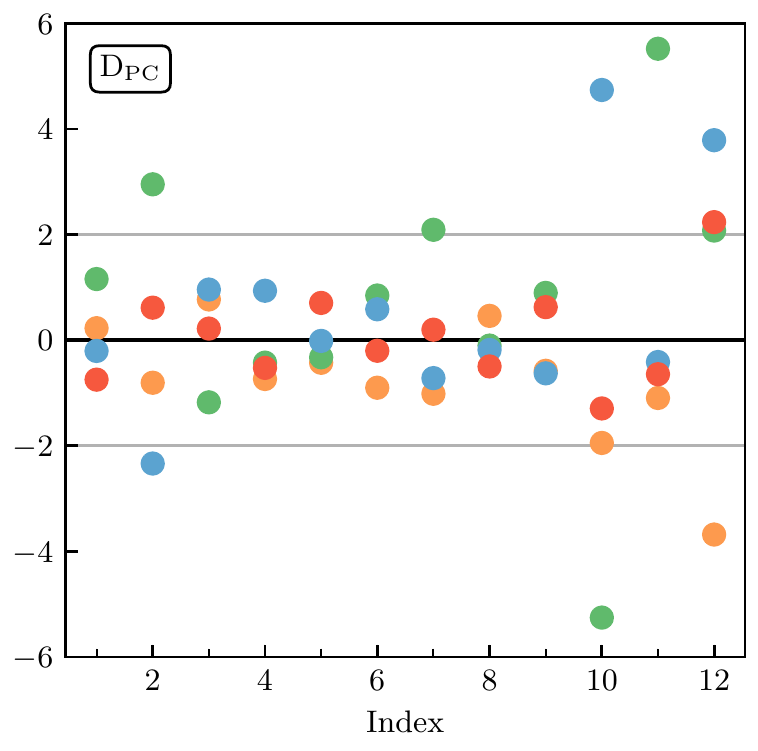}
}
\caption{Differential cross section coefficients from EKM $R=0.9$ potential evaluated at $\Elab = 150$\,MeV.
In (a), the training data are denoted by dots, whereas the validation data locations are indicated by the minor ticks.
See Fig.~\ref{fig:toy_coeffs_and_diagnostics} for the remaining figure notation.
The problems with the diagnostics for $c_3$ and $c_4$ are driven primarily by the backwards angles, which are unusually large.
Outlier values at large index in the pivoted Cholesky plot point to problems with non-stationarity.
}
\label{fig:dsg_coeff_analysis}
\end{figure*}

\begin{figure*}
\captionsetup[subfloat]{margin=0pt}
\subfloat[]{%
\includegraphics{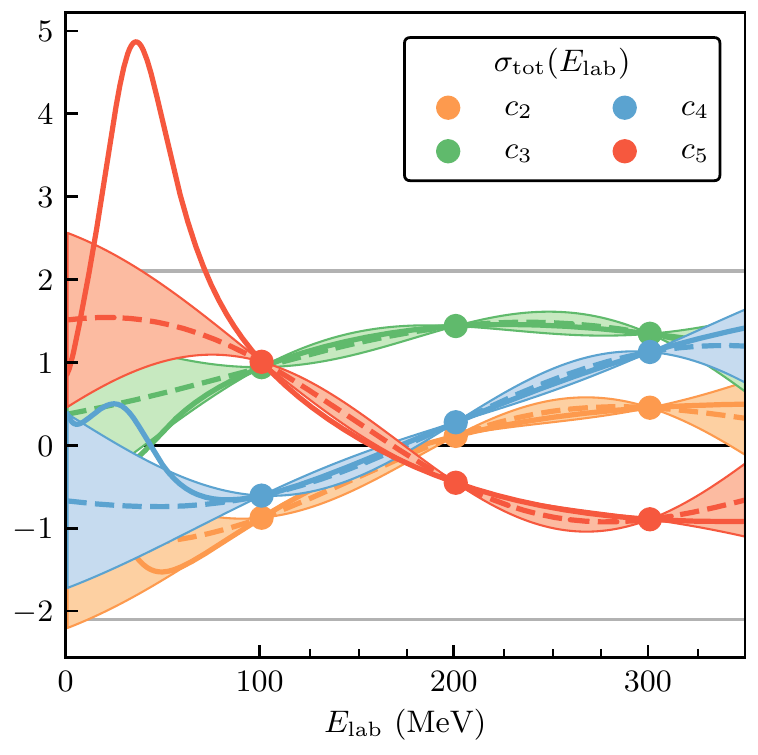}
}
\subfloat[]{%
\includegraphics{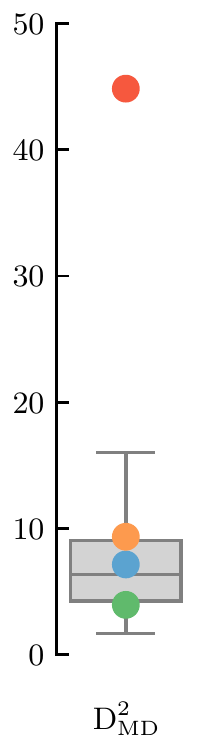}
}
\subfloat[]{%
\includegraphics{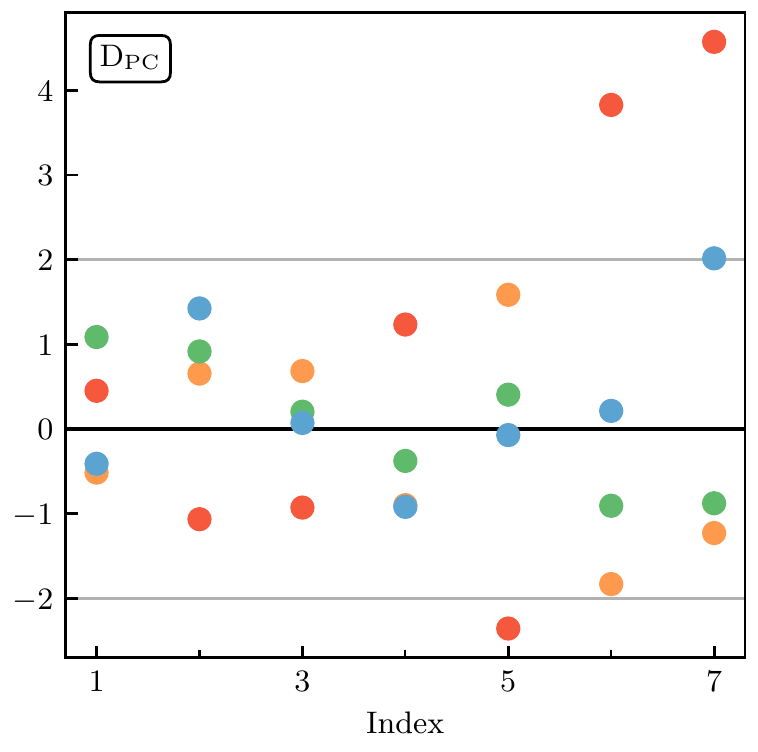}
}
\caption{
(a) Total cross section coefficients using the EKM $R=0.9$ potential, as shown in Ref.~\cite{MelendezBayesiantruncationerrors2017}.
The training data are denoted by dots, whereas the validation data locations are indicated by the minor ticks.
See Fig.~\ref{fig:toy_coeffs_and_diagnostics} for the remaining figure notation.
(b) The $\DMD^2$ clearly shows a problem with $c_5$, whereas the others are behaving well.
(c) When decomposed as $\DVAR{PC}$, the issue with $c_5$ appears at large index, indicating potential problems with non-stationarity.
If low $\Elab$ validation data were included, the problem becomes much more visible.
}
\label{fig:total_cross_coeff_analysis}
\end{figure*}

\begin{figure*}
\captionsetup[subfloat]{margin=0pt}
\subfloat[]{%
\includegraphics{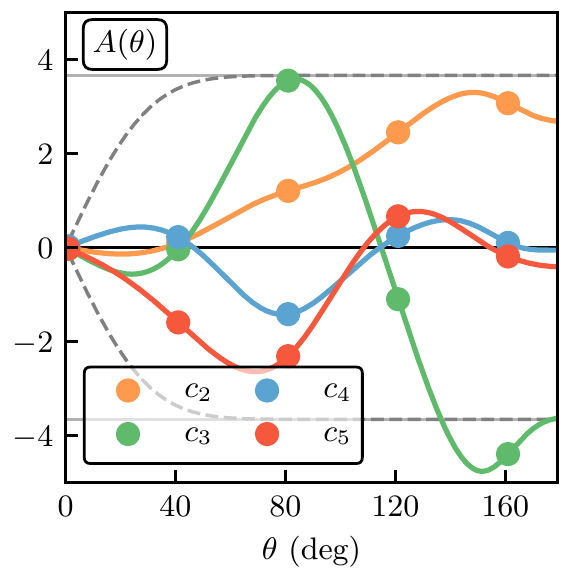}
\label{fig:spin_obs_coeffs_and_truncation_coeffs}
}
\subfloat[]{%
\includegraphics{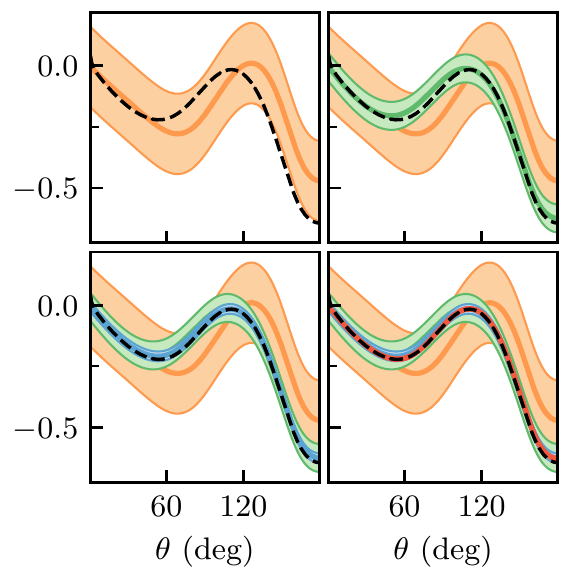}
\label{fig:spin_obs_coeffs_and_truncation_unc}
}
\subfloat[]{%
\includegraphics{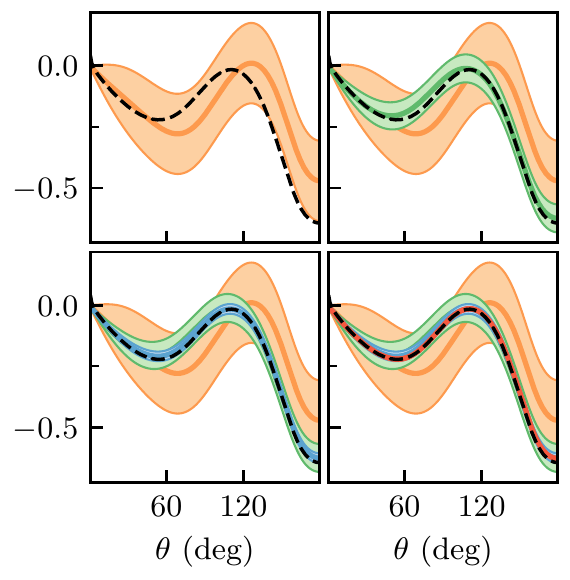}
\label{fig:spin_obs_coeffs_and_truncation_constrained}
}
\caption{The spin observable $A$, calculated at $\Elab=96$\,MeV using the EKM $R=0.9$\,fm potential, with the magnetic-moment interaction neglected for simplicity. $A$ is then constrained to be zero at $\theta=0^\degr$.
(a) The observable coefficients and the points used to learn $\sdth$ and $\ell$ of their underlying distribution.
One can choose to estimate using the underlying distribution, shown as a gray line, or the underlying distribution conditioned to be 0 at $\theta=0^\degr$, shown as a darker gray dashed line.
The conditioned distribution more closely follows the pattern of the coefficients, and will more accurately describe the higher-order coefficients due to the symmetry of the system.
(b) The EFT predictions and their $2\sigma$ truncation errors using the underlying unconstrained coefficient distribution. The truncation error does not vanish at $\theta=0^\degr$.
(c) The EFT predictions and their $2\sigma$ truncation errors using the constrained coefficient distribution. Now the truncation error does vanish at $\theta=0^\degr$, as expected by symmetry arguments.
}
\label{fig:spin_obs_coeffs_and_truncation}
\end{figure*}

Let us begin by testing the convergence pattern of the EFT for the total and differential cross sections.
To do so, we follow the path set out in Sec.~\ref{sec:toy_application}, that is, we extract the observable coefficients using the physically motivated assumptions of $\genobsref$ and $Q$ described above, and evaluate their features using the diagnostics of Sec.~\ref{sec:model_checking}.
The analysis of the differential cross section evaluated at $\Elab = 150$\,MeV is given in Fig.~\ref{fig:dsg_coeff_analysis}.
The coefficients look beautifully \iid\ and stationary until $\theta$ approaches backward angles, at which point $c_3$ and $c_4$ grow in size.
This observation is reflected in the diagnostics computed at the validation points.
It is clear that both $c_3$ and $c_4$ are outliers by the Mahalanobis distance, whereas $c_2$ and $c_5$ are appropriately sized.
The pivoted Cholesky provides some insight into the problem: unusually sized values at high index point to problems in the correlation structure or non-stationarity.
And it turns out that these outliers at large index are exactly the validation points at large $\theta$, as can be verified by removing all points greater than $\theta = 125^\degr$ from the diagnostic.
If the backward angles are excluded, both diagnostics show excellent agreement with our model assumptions.

Now we turn to the analysis of the total cross section, which was extensively studied in Refs.~\cite{furnstahl_quantifying_2015,MelendezBayesiantruncationerrors2017} for differently regulated potentials.
The coefficients and their diagnostics are shown in Fig.~\ref{fig:total_cross_coeff_analysis}.
The choice of test and validation points merit mention here.
The highly non-stationary behavior at low energy is not representative of the high-energy data and would bias the analysis if a stationary process were used.
This is a standard artifact that appears in our tested $NN$ observables when plotted against lab energy~\cite{MelendezBayesiantruncationerrors2017}.
Hence these points were omitted from this preliminary analysis.
Additionally, the training and test points must be spaced out---due to the very large length scale in this system---otherwise numerical instabilities appear.
Given these caveats, the analysis of the total cross section nevertheless show interesting patterns.
The diagnostics show that $c_5$ is a clear outlier, even when restricted to high energy data.
The pivoted Cholesky points again to the onset of non-stationarity or a misestimation of the length scale.

How does chiral EFT fare when subjected to these diagnostics?
Both the total and differential cross sections show behavior that is consistent with our convergence model in some regimes, yet inconsistent in others.
In this case, there are plausible physical explanations for the diagnostics of unusual size.
Likely explanations for the failures of the differential and total cross sections are a misspecification of $Q$ at backward angles and low energies, respectively. This problem with the choice of $Q$ will then result in incorrect truncation error estimates.
It may be that the correct expansion parameter takes into account the momentum transfer for angle-dependent observables like the differential cross section.
Meanwhile, the non-stationarity appearing at low energies occurs in the regime where $\prel \approx m_\pi$, pointing to the possibility that the crossover region of $Q$ was not parameterized correctly.
Barring the wrong choice of $Q$, these diagnostic failure modes could simply point to the possibility that the EFT convergence pattern fails in these regimes according to the definition of convergence proposed in this work.

We have shown how observable coefficients can be analyzed in real EFT predictions; truncation errors follow as in Sec.~\ref{sec:toy_application}.
We illustrate this for the spin observable $A$, which is constrained to be zero at $\theta = 0^\degr$ if the (small) magnetic-moment interaction is neglected, as it is here.
Therefore contributions from all orders of the EFT are zero as well.
We can accommodate the constraint using the formalism described in Sec.~\ref{sub:application_types} and shown in Fig.~\subref*{fig:gp_applications_constraint}.
Two sets of EFT truncation errors are proposed in Fig.~\ref{fig:spin_obs_coeffs_and_truncation}, without and with the constraint.
The bands in Fig.~\subref*{fig:spin_obs_coeffs_and_truncation_unc} are unrealistically large at $\theta=0^\degr$ given the information we know about this system.
On the other hand, the bands in Fig.~\subref*{fig:spin_obs_coeffs_and_truncation_constrained} vanish at $\theta=0^\degr$ as expected. This example uses $\Elab=96$\,MeV, but the point is general.
The truncation errors are smoothly connected between the constrained and unconstrained regions in a manner harmonious with the way that lower-order coefficients approach zero. That is, 
the extent of the domain affected by this constraint is dependent on the length scale $\ell$ extracted from the known coefficients.

%%%%%%%%%%%%%%%%%%%%%%%%%%%%%%%%%%%%%%%%%%%%%%%%%%%%%%%%%%%%%%%

\section{Summary and Outlook}
\label{sec:summary}

Order-by-order predictions of a well-behaved EFT should converge
regularly toward the all-orders value.
We have formalized this idea into a falsifiable EFT convergence model.
Rigorous estimates of the truncation error---induced by stopping the EFT expansion
at a finite order---are now made possible, as are novel techniques for estimating EFT-related quantities, such as the expansion parameter $Q$, breakdown scale $\Lambda_b$, and low-energy constants (LECs).
We believe that formalizing and testing convergence is a crucial step when using an EFT, otherwise failure may go unrecognized.

This work answers several open questions faced by EFT practitioners that were left unresolved by our earlier work~\cite{furnstahl_quantifying_2015,MelendezBayesiantruncationerrors2017,WesolowskiExploringBayesianparameter2019}.

(1) \emph{EFTs make predictions $\genobs_n(\kinparvec)$ that may be correlated in $\kinparvec$, which in turn induces correlations in the truncation errors; how should that be taken into account when making predictions and fitting?}
With the EFT power counting as our guide, we propose a change of variables that allows us to isolate the correlated quantities: observable coefficients $c_n(\kinparvec)$ (see Fig.~\ref{fig:preds_to_coeffs}).
The truncation error distribution follows by modeling the $c_n$ as draws from a Gaussian process (GP) and summing all higher order contributions.
Originally, we had treated predictions at each kinematic value in a pointwise manner~\cite{furnstahl_quantifying_2015,MelendezBayesiantruncationerrors2017} and later modeled correlations in two extreme limits, which correspond to zero and infinite GP length scale $\ell$~\cite{WesolowskiExploringBayesianparameter2019}.
Now $\ell$ can be estimated using data across multiple curves.
This GP model could be adapted to correlations between discrete predictions as well.

(2) \emph{Is this Bayesian formalism computationally demanding?}
Bayesian updating of our convergence model parameters $\param$ is made simple and analytic by the use of conjugate priors.
Our self-contained treatment shows how prior information on naturalness and correlations is updated using \emph{all} available data, and derives a novel analytic posterior for the EFT expansion parameter $Q$ (see Fig.~\ref{fig:toy_Q_ell_posterior} and Appendix~\ref{app:derivations}).

(3) \emph{What about computationally expensive systems?}
If it is too expensive to compute an EFT prediction at all points $\kinparvec$ or orders $n$, then $\param$ and $Q$, which control the truncation error distribution, can still be estimated from the available predictions without problem.
(Even for inexpensive systems, if some low-order corrections are known to be unrepresentative of those at higher orders, one would not want to use these in estimating the truncation error; for example, $c_1$ vanishes by symmetry in chiral EFT.)
In these cases, the posteriors of $\param$ and $Q$ are simply more influenced by the priors.
After estimating $\param$ and $Q$, our convergence model then seamlessly combines truncation estimates with fast interpolation formulae to make predictions at all desired $\kinparvec$ (see Fig.~\ref{fig:gp_applications} and Sec.~\ref{sub:application_types}, which also shows how to incorporate symmetry constraints).

(4) \emph{How do these results affect EFT fitting?}
We derived a novel likelihood function [Eq.~\eqref{eq:one_obs_parameter_estimation}] that enables the EFT fit to account for all higher-order terms and their correlations in $\kinparvec$. This extends to fitting with interpolation as well, allowing more experimental data to be used than might otherwise be computationally feasible.
Because it is Gaussian, this likelihood can easily be incorporated into existing fitting procedures.

(5) \emph{How is this falsifiable?}
We provide a menu of model checking diagnostics that enable anyone to assess whether an EFT is working as advertised.
We emphasize that these diagnostics require reference distributions (whose functional forms are given in Sec.~\ref{sub:emulator_choices}) to assign statistical significance to their output.
Claims about EFT convergence or truncation errors can then be falsified at any desired level of significance.
We demonstrate the usefulness of these diagnostics on a toy example in Sec.~\ref{sec:toy_application}, which exemplifies the general workflow of testing an EFT\@.
Section~\ref{sec:NN_scattering} does the same for selected $NN$ scattering observables in chiral EFT\@.
The success and failure modes within chiral EFT point to new research directions to pursue.
This ability to probe for failure using statistical diagnostics
and visualization means that our framework is not just a way to establish theoretical
error bars, but provides powerful tools for identifying limitations in our statistical model \emph{and} in the physical understanding encoded in an EFT\@.

The stage is now set for a wide range of applications within chiral EFT; some of the paths
in progress are
\begin{itemize}
  \item A reevaluation of the EFTs critiqued with the pointwise convergence model in Ref.~\cite{MelendezBayesiantruncationerrors2017} in light of the GP model. Additionally, new variants of chiral EFT, which may show promising convergence properties in the $NN$ sector and fewer regulator artifacts~\cite{ReinertSemilocalmomentumspaceregularized2018,EntemHighqualitytwonucleonpotentials2017}, should be included in the analysis.
  \item A systematic study of EFT convergence beyond $NN$ observables to assess its validity.
  This includes few-body, nuclear matter, and Compton scattering observables.
  \item An exploration of the effects of our proposed likelihood [Eq.~\eqref{eq:one_obs_parameter_estimation}] on the LECs' posteriors, and in particular the effect on three-body LECs.
\end{itemize}
The models and tools we have developed apply generally to other EFTs and perturbative theories in general, so we anticipate
many other applications.
The pointwise and curvewise models and associated model checking diagnostics are implemented in 
the \texttt{gsum}~\cite{gsum} Python package hosted at \href{https://github.com/buqeye/gsum}{github.com/buqeye/gsum} and can be freely downloaded and used.
This includes a notebook that creates all figures shown here, among other examples.

%%%%%%%%%%%%%%%%%%%%%%%%%%%%%%%%%%%%%%%%%%%%%%%%%%%%%%%%%%%%%%%

\acknowledgments{We thank Derek Bingham for useful discussions.
JM thanks Akira Horiguchi and John Honaker for contributions to an early draft of this project. We thank Natalie Klco for contributions to the early stages of, and discussions on, this work. 
% Useful feedback on the manuscript was provided by ??.
The work of RJF and JAM was supported in part by the National Science Foundation
under Grant No.~PHY--1614460 and the NUCLEI SciDAC Collaboration under
US Department of Energy MSU subcontract RC107839-OSU\@.
The work of JAM was also supported in part by the U.S. Department of Energy, Office of Science, Office of Workforce Development for Teachers and Scientists, Office of Science Graduate Student Research (SCGSR) program. The SCGSR program is administered by the Oak Ridge Institute for Science and Education for the DOE under contract number DE‐SC0014664\@.
The work of DRP was supported by the US Department of Energy under
contract DE-FG02-93ER-40756 and by the 
 ExtreMe Matter Institute EMMI
at the GSI Helmholtzzentrum f\"ur Schwerionenphysik, Darmstadt, Germany.
The work of SW was supported in part by Salisbury University.
}

%%%%%%%%%%%%%%%%%%%%%%%%%%%%%%%%%%%%%%%%%%%%%%%%%%%%%%%%%%%%%%%

\appendix

%%%%%%%%%%%%%%%%%%%%%%%%%%%%%%%%%%%%%%%%%%%%%%%%%%%%%%%%%%%%%%%

\section{Derivations}
\label{app:derivations}

The mathematical details of the model described in Sec.~\ref{sec:the_model} are shown here.
Due to the choice of conjugate priors on $\mu,\sdth^2$, almost all posteriors of interest can be provided in closed form.

Before getting into the fine details of the GP model discussed in this work, we believe it instructive to begin with a short aside.
We have considered a simpler pointwise model in previous works~\cite{furnstahl_quantifying_2015,MelendezBayesiantruncationerrors2017}, which is shown graphically in Fig.~\ref{fig:random_annotated_curves}.
To our knowledge, no one has yet worked out analytic expressions for the case of conjugate priors, whose derivations mirror the correlated case.
Hence we will provide the results below for completeness.
Again, conjugacy allows for efficient computation of truncation errors, parameter posteriors, and evidences using standard statistical libraries.
We promote the use of these conjugate priors over the options laid out in previous works.
% We do not find the form of the priors to be unnecessarily restrictive and hence we promote its use.

See Table~\ref{tab:notation} for a summary of the vector notation.
Footnote~\ref{statnote} gives a brief overview of the statistics notation.

\renewcommand{\tabcolsep}{3pt}

\begin{table}[tb]
\caption{Notation for vectors in different spaces.}
\label{tab:notation}
% \renewcommand{\arraystretch}{1.4}%
% \setlength\extrarowheight{1pt}
% Use S prefix to enable the cellspace package
% which makes the spacing look nice.
\begin{ruledtabular}
\begin{tabular}{ScScScSc}
  Notation & Example & Description \\
  \colrule
  --- & $\kinparvec\in\mathbb{R}^d$ & Input space (suppressed) \\
  Bold & $\inputvec{c}_n\in\mathbb{R}^N$ & A set using inputs $\{\kinparvec_i\}_{i=1}^N$ \\
  Arrow & $\ckvec \in\mathbb{R}^{\nc}$ & Orders $c_0, c_1, \dots, c_k$ \\
  Bold Italics & $\param$ & GP parameters $\{\mu, \sdth, \ell\}$ \\
  --- & $\mu$, $\hypm$, $\hypdisp$ & Mean basis space \\
  Composition & $\ordervec{\inputvec{c}}_k \in \mathbb{R}^{\nc} \times \mathbb{R}^{N}$ & Combination of above \\
      & $\kinparvecset \in \mathbb{R}^N \times \mathbb{R}^d$ &
\end{tabular}
\end{ruledtabular}
\end{table}

\subsection{Pointwise Model}
\label{sub:pointwise_model}

Consider a set of EFT predictions $\{\genobs_n\}$, each of which is a number rather than a function.
In this case, the quantity of interest $\genobs_n$ could itself be a scalar or we could be considering a functional quantity $\genobs_n(\kinparvec)$ at a specific input point $\kinparvec$ in its domain.
Here we wish to predict the truncation error independently of the function values at any surrounding input points.
This procedure could be repeated at all points in the domain to produce error bands, as was done in Ref.~\cite{MelendezBayesiantruncationerrors2017}.
This is what we call the \emph{pointwise} model.

If the expansion parameter $Q$ and reference scale $\genobsref$ are assumed to be known, then the data $\genobsvec$ can be converted to coefficients $\ckvec$,
\begin{align}
  \genobsvec \equiv
  \transpose{
  \begin{bmatrix}
    \genobs_0 & \genobs_1 & \cdots & \genobs_k
  \end{bmatrix}
  }
  ~\Rightarrow~
  \ckvec \equiv
  \transpose{
  \begin{bmatrix}
    c_0 & c_1 & \cdots & c_k
  \end{bmatrix}
  }.
\end{align}
Here we assume that the coefficients $c_n$ are normal given $\sdth^2$ and that $\sdth^2$ has a scaled inverse-$\chi^2$ prior~\cite{GelmanBayesianDataAnalysis2013},
\begin{align}
  c_n \given \sdth^2 & \overset{\text{\tiny iid}}{\sim} \normal(0, \sdth^2) \label{eq:cn_prior_uncorr} \\
  \sdth^2 & \sim \invchisq(\hypprior\hypdf, \hypprior\hyptau^2). \label{eq:sigma_prior_uncorr}
\end{align}
Equation~\eqref{eq:sigma_prior_uncorr} has qualitatively similar properties to priors used in past works~\cite{furnstahl_quantifying_2015,MelendezBayesiantruncationerrors2017}: it is strictly positive, can penalize $\sdth^2$ for being or large or too small, and reduces to a scale-invariant prior for $\hypprior\hypdf = 0$.
Unlike past priors on $\sdth^2$, the scaled inverse-$\chi^2$ permits intuitive analytic posteriors for $\delta\genobs_k$ and $Q$ for \emph{all} choices of $\hypprior\hypdf$ and $\hypprior\hyptau^2$.

Analogously to Eq.~\eqref{eq:discr_k_prior}, one can derive from Eqs.~\eqref{eq:discrepancy_k_definition}, \eqref{eq:normal_sum}, and \eqref{eq:cn_prior_uncorr} that
\begin{align} \label{eq:discr_k_prior_uncorr}
    \delta \genobs_k \given \sdth^2, Q \sim \normal{\left[0, \genobsref^2\frac{Q^{2(k+1)}}{1 - Q^2}\sdth^2\right]}.
\end{align}
The above equations outline our prior beliefs about the coefficients, their widths, and how they combine to form a discrepancy term for $\genobs_k$.
If $\sdth$ was precisely known from \emph{a priori} arguments or otherwise, then Eq.~\eqref{eq:discr_k_prior_uncorr} is sufficient to estimate the truncation error.
Since this is not the case in general, we would like to update these priors based on order-by-order predictions.

Our belief about an unseen coefficient $c_n$ is updated via
\begin{align} \label{eq:cn_posterior_predictive_integral}
  \pr(c_n \given \ckvec) & = \int_0^\infty \pr(c_n \given \sdth^2) \pr(\sdth^2 \given \ckvec) \dd{\sdth^2}.
\end{align}
By marginalizing in information about $\sdth^2$, it is clear that the effect of data on $c_n$ flows through $\sdth^2$.
To compute the \emph{posterior predictive distribution}, Eq.~\eqref{eq:cn_posterior_predictive_integral}, we must first understand how the observed coefficients $\ckvec$ impact our understanding of $\sdth^2$.
This is where the utility of conjugate priors is realized.
The posterior of $\sdth^2$ has the same functional form as the prior---an inverse $\chi^2$ distribution---but with $\cond\hypdf$ and $\cond\hyptau^2$ that are updated by the data.
This property simplifies our analysis greatly: computing the posterior $\sdth^2 \given \ckvec$ is reduced to determining $\cond\hypdf$ and $\cond\hyptau^2$.
From Bayes' theorem and the fact that the $\ckvec$ are marginally independent given $\sdth^2$,
\begin{align}
  \pr(\sdth^2 \given \ckvec) & \propto \pr(\ckvec \given \sdth^2) \pr(\sdth^2) = \pr(\sdth^2) \prod_n \pr(c_n \given \sdth^2) \notag \\
  & \propto \frac{1}{\sdth^{\hypprior\hypdf+\nc+2}} \exp[-\frac{1}{2\sdth^2}\left(\hypprior\hypdf\hypprior\hyptau^2 + \ckvecsq\right)].
\end{align}
%Above we have defined $\nc$ to be the number of non-zero coefficients in $\ckvec$ 
Above we have defined $\nc$ to be the number of coefficients in $\ckvec$ that are useful for induction 
(which is not always $k+1$, see Ref.~\cite{furnstahl_quantifying_2015}).
%Above we have defined the number of coefficients $\nc \equiv |\ckvec|$.
Upon comparison with Eq.~\eqref{eq:invchi2_density}, we can read off
\begin{align}
  \cond\hypdf & = \hypprior\hypdf + \nc \label{eq:posterior_df} \\
  \cond\hypdf\cond\hyptau^2 & = \hypprior\hypdf\hypprior\hyptau^2 + \ckvecsq. \label{eq:posterior_tau}
\end{align}
Therefore, we have shown that
\begin{align} \label{eq:marginal_variance_posterior}
    \sdth^2 \given \ckvec \sim \invchisq(\cond\hypdf, \cond\hyptau^2),
\end{align}
with $\cond\hypdf$ and $\cond\hyptau^2$ defined by Eqs.~\eqref{eq:posterior_df} and~\eqref{eq:posterior_tau}.

Now that the posterior for $\sdth^2$ has been established as Eq.~\eqref{eq:marginal_variance_posterior}, we can return our attention to Eq.~\eqref{eq:cn_posterior_predictive_integral}.
This integral evaluates to
\begin{align} \label{eq:cn_t_pdf}
  \pr(c_n \given \ckvec) = \frac{1}{\sqrt{\pi \cond\hypdf\cond\hyptau^2}} \frac{\Gamma(\frac{\cond\hypdf+1}{2})}{\Gamma(\frac{\cond\hypdf}{2})} \left(1 + \frac{c_n^2}{\cond\hypdf\cond\hyptau^2}\right)^{-\frac{\cond\hypdf+1}{2}}.
\end{align}
Equation~\eqref{eq:cn_t_pdf} is the pdf for the Student-$t$ distribution with degrees of freedom $\cond\hypdf$, mean $0$, and scale $\cond\hyptau$, i.e.,
\begin{align}
    c_n \given \ckvec & \sim t_{\cond\hypdf}(0, \cond\hyptau^2). \label{eq:cn_posterior_uncorr}
\end{align}
In contrast to the notation of the Gaussian distribution, the variance of the Student-$t$ is given by $\cond\hypdf\cond\hyptau^2/(\cond\hypdf-2)$ and is only defined when $\cond\hypdf > 2$.

Let us pause and ponder the implications of what we have derived.
Equations~\eqref{eq:cn_prior_uncorr}, \eqref{eq:posterior_df}--\eqref{eq:marginal_variance_posterior}, and \eqref{eq:cn_posterior_uncorr} fit together within one cohesive picture and relate to common estimators employed in frequentist statistics.
\begin{itemize}
    \item The variance of $\sdth^2$ decreases as the degrees of freedom $\cond\hypdf$ increases.
    This uncertainty in $\sdth^2$ maps directly to the heaviness of the tails in the $t$ distribution, which is also controlled by $\cond\hypdf$.
    \item A strict prior ($\hypprior\hypdf \gg 1$) or a lot of data ($\nc \gg 1$) both lead to $\pr(\sdth^2 \given \ckvec)$ being sharply peaked at $\sdth^2 \approx \cond\hyptau^2$.
    This informative distribution for $\sdth^2$ corresponds to a $\pr(c_n \given \ckvec)$ that approximates a Gaussian with variance $\cond\hyptau^2$.
    \item Given a lot of data, then $\sdth^2 \approx \cond\hyptau^2 \approx \ckvecsq/\nc$, which is a frequentist estimator known as the \emph{sample variance}.
    In this case $\pr(c_n \given \ckvec)$ will approximate a Gaussian with $\ckvecsq/\nc$ as its variance.
    \item Rather than integrating over $\sdth^2$, we could use its mean value as a point estimate for the variance in Eq.~\eqref{eq:cn_prior_uncorr}.
    The mean of Eq.~\eqref{eq:marginal_variance_posterior} is given by $\cond\hypdf\cond\hyptau^2/(\cond\hypdf-2)$, which is exactly the variance of the $t$ distribution in Eq.~\eqref{eq:cn_posterior_uncorr}.
\end{itemize}
By computing or assuming values for $\cond\hypdf$ and $\cond\hyptau^2$, we are able to relax the assumption of a known $\sdth^2$ in Eq.~\eqref{eq:cn_prior_uncorr} by giving $c_n$ a distribution with tails and a scale compatible with our understanding of $\sdth^2$.
A Bayesian perspective allows us to incorporate prior beliefs about naturalness and update them with finite sets of data in a manner that is harmonious with the limiting cases assumed in frequentist statistics.

A better understanding of unseen observable coefficients $c_n$ is useful as a means to better understand the truncation error $\delta\genobs_k$.
The steps in deriving Eq.~\eqref{eq:cn_posterior_uncorr} lead analogously to the posterior
\begin{align}
  \delta\genobs_k \given \genobsvec, Q & \sim t_{\cond\hypdf}{\left(0, \genobsref^2\frac{Q^{2(k+1)}}{1 - Q^2}\cond\hyptau^2\right)}, \label{eq:genobsk_posterior_uncorr}
\end{align}
where $\cond\hypdf$ and $\cond\hyptau^2$ are computed via Eqs.~\eqref{eq:posterior_df} and~\eqref{eq:posterior_tau}.
The distribution for the full prediction $\genobs = \genobs_k + \delta\genobs_k$ is thus
\begin{align} \label{eq:full_obs_dist_uncorr}
  \genobs \given \genobsvec, Q & \sim t_{\cond{\hypdf}}{\left(\genobs_k, \genobsref^2\frac{Q^{2(k+1)}}{1 - Q^2}\cond\hyptau^2\right)}.
\end{align}
Equation~\eqref{eq:full_obs_dist_uncorr} is a novel result from this work and is the most straightforward method to compute posteriors of $\genobs$ in the pointwise model.
Statistical packages such as SciPy have built-in $t$ distributions which can easily calculate degree of belief intervals for Eq.~\eqref{eq:full_obs_dist_uncorr}.

Until now, we have assumed that the expansion parameter $Q$ is known, but in practice this may not be the case.
In fact, $Q$ may even be a function of $\kinparvec$ in certain circumstances, which we assume here for generality.
Fortunately, conjugacy permits the analytic computation of the (unnormalized) posterior
\begin{align}
    \pr(\inputvec{Q} \given \genobsvecset) \propto \pr(\genobsvecset \given \inputvec{Q}) \pr(\inputvec{Q}).
\end{align}
The right-hand side is (up to a simple integral over $Q$) the model evidence, which is a useful quantity for model comparison,
and the likelihood $\pr(\genobsvecset \given \inputvec{Q})$ is exactly calculable.
The pointwise model assumes that each of the $\genobsvec(\kinparvec_i)$ are independent and each have their own $\sdth_i$, so that $\pr(\genobsvecset \given \inputvec{Q}) = \prod_i \pr[\genobsvec(\kinparvec_i) \given Q(\kinparvec_i)]$.
We consider each $\kinparvec_i$ individually and drop $\kinparvec_i$ for convenience.
Note that since $\genobs_0 = \genobsref c_0$ and $\genobs_n = \genobs_{n-1} + \genobsref c_n Q^n$, we can make a change of variables
\begin{align} \label{eq:d_given_q_uncorr}
  \pr(\genobsvec \given Q) = \frac{\pr(\ckvec)}{\prod_n \abs{\genobsref Q^n}},
\end{align}
where $c_0 = \genobs_0 / \genobsref$ and $c_n = \Delta\genobs_n / \genobsref Q^n$ implicitly depend on $Q$.
Thus we need only compute the joint distribution $\pr(\ckvec)$.
Although the distribution of one coefficient $c_n$ follows a $t$ distribution Eq.~\eqref{eq:cn_posterior_uncorr}, the same is not true of the joint distribution of multiple coefficients $\ckvec$.
Nevertheless, we can find an analytic form for their distribution using the clever manipulation of normalization constants.
Let primes denote unnormalized distributions.
Then both
\begin{align}
  \pr(\sdth^2 \given \ckvec) = \frac{(\hypprior\hypdf\hypprior\hyptau^2/2)^{\hypprior\hypdf/2}}{\pr(\ckvec)\Gamma(\hypprior\hypdf/2)} \invchisq'(\hypprior\hypdf, \hypprior\hyptau^2) \prod_n \frac{1}{\sqrt{2\pi}} \normal'(0, \sdth^2),
\end{align}
from Bayes' theorem, and
\begin{align}
  \pr(\sdth^2 \given \ckvec) = \frac{(\cond\hypdf\cond\hyptau^2/2)^{\cond\hypdf/2}}{\Gamma(\cond\hypdf/2)} \invchisq'(\cond\hypdf, \cond\hyptau^2),
\end{align}
from Eq.~\eqref{eq:marginal_variance_posterior}, are true statements.
Since $\pr(\ckvec)$ does not depend on $\sdth^2$, equating the normalization coefficients yields the desired quantity
\begin{align} \label{eq:pr_ckvec_uncorr}
  \pr(\ckvec) = \frac{\Gamma(\cond\hypdf/2)}{\Gamma(\hypprior\hypdf/2)} \sqrt{\frac{1}{(2\pi)^{\nc}} \frac{(\hypprior\hypdf\hypprior\hyptau^2/2)^{\hypprior\hypdf}}{(\cond\hypdf\cond\hyptau^2/2)^{\cond\hypdf}}}.
\end{align}
Through Eqs.~\eqref{eq:d_given_q_uncorr} and~\eqref{eq:pr_ckvec_uncorr} we have completely specified $\genobsvec \given Q$, and $\genobsvecset \given \inputvec{Q}$ is simply the product of these quantities.

Finally, we can write an unnormalized expression for the expansion parameter posterior by dropping terms independent of $Q$ and multiplying $\genobsvec(\kinparvec_i) \given Q(\kinparvec_i)$ at each $\kinparvec_i$
\begin{align}
  \pr(\inputvec{Q} \given \genobsvecset) \propto \pr(\inputvec{Q}) \prod_i \left[\cond\hyptau_i^{\cond\hypdf_i} \prod_n Q_i^n\right]^{-1}.
\end{align}
Rather than dealing with the multidimensional object $\inputvec{Q}$, one can instead parameterize $Q$ as a low-energy scale $f(\kinparvec)$ and a high-energy scale $\Lambda_b$, known as the breakdown scale of the EFT\@.
These are related to $Q$ via $Q(\kinparvec) = f(\kinparvec)/\Lambda_b$.
The posterior for $\Lambda_b$, assuming $f(\kinparvec)$ is known, has a simple relationship to the posterior for $Q$:
\begin{align}
    \pr(\Lambda_b \given \genobsvecset, \inputvec{f}) & \propto \pr(\genobsvecset \given \inputvec{f}, \Lambda_b) \pr(\Lambda_b) \notag \\
    & = \pr(\genobsvecset \given \inputvec{Q}) \pr(\Lambda_b) \notag \\
    & \propto \pr(\Lambda_b) \prod_i \left[\cond\hyptau_i^{\cond\hypdf_i} \prod_n Q_i^n\right]^{-1}. \label{eq:Lb_uncorr_posterior}
\end{align}
This is a more general version of the posterior given in Ref.~\cite{MelendezBayesiantruncationerrors2017}, where the restriction $\hypprior\hypdf=0$ was required.
Here, thanks to conjugacy, $\hypprior\hypdf$ and $\hypprior\hyptau^2$ are free to be chosen by the modeler.
The extension to further parameters of $Q$ requires only a modification of the prior $\pr(\Lambda_b)$ in Eq.~\eqref{eq:Lb_uncorr_posterior}.

\subsection{Gaussian Process Model}
\label{sub:gaussian_process_model}

Here we will expand on the template provided by Appendix~\ref{sub:pointwise_model} and derive analogous results for the GP truncation error model introduced in Sec.~\ref{sec:the_model}.
The results in this Appendix follow exactly from the analysis of the Bayesian linear model (see Ref.~\cite{o1994bayesian}, chaps.\ 9 and 10) but are reproduced here for completeness.
As a reminder, we have assumed that the $c_n$ are GPs with the following form:
\begin{align}
  c_n(\kinparvec) \given \param \overset{\text{\tiny iid}}{\sim} \GP[\basisfunc^\trans(\kinparvec)\mu, \sdth^2 r(\kinparvec,\kinparvec';\ell)], \label{eq:cn_iid_with_basis}
\end{align}
and placed a normal-inverse-$\chi^2$ prior on $\mu,\sdth^2$,
\begin{align*}
  \mu,\sdth^2 \sim \ninvchisq(\hypprior\hypm, \hypprior\hypdisp, \hypprior\hypdf, \hypprior\hyptau^2). \eqrevisited{eq:ninvchi2_prior}
\end{align*}
We have generalized the parametrization as promised in Sec.~\ref{sub:gp_truncation} by introducing a length $p$ vector of functions $\basisfunc(\kinparvec)$ that multiplies a length $p$ vector of regression coefficients $\mu$ to allow for a non-constant mean function as in, e.g., Ref.~\cite{BastosDiagnosticsGaussianProcess2009}.  Both $\hypprior\hypm$ and $\hypprior \hypdisp$, previously scalars, are promoted to a length $p$ vector and a $p \times p$ matrix, respectively.
For the examples of EFT truncations shown in this work, we use $\basisfunc(\kinparvec) = 1$.

As before, we would like to condition on data to gain a better understanding of the observable coefficients and ultimately the discrepancy $\delta\genobs_k(\kinparvec)$.
Again, assume that $Q$ and $\genobsref$ are known 
so that our data are $N$ vectors of coefficients $\ckvec$ (each of length $\nc$) at the chosen input points. The combined ($\nc\times N$)-shaped  observations are denoted $\ckvecset$ (see Table~\ref{tab:notation}).
%that our data are $\nc\times N$ shaped observations %$\ckvecset = \{c_n(\kinparvecset)\}_n$.
The posterior predictive distribution for a new curve $c_n(\kinparvec)$ is then
\begin{align} \label{eq:cn_gp_posterior_predictive_integral}
    \pr(c_n(\kinparvec) \given \ckvecset) = \int \pr[c_n(\kinparvec) \given \param] \pr(\param \given \ckvecset) \dd{\param},
\end{align}
where $\param = \{\mu, \sdth^2, \ell\}$.
As before, the data influence $c_n(\kinparvec)$ by updating our beliefs about $\param$.
Because of the conjugate prior on $\mu$ and $\sdth^2$, they can be updated and integrated out of Eq.~\eqref{eq:cn_gp_posterior_predictive_integral} analytically, but there is no conjugate prior for $\ell$.
For now, assume that $\ell$ is known; we will return to finding its posterior later.
With a posterior for $\ell$, one could use its MAP value as a point estimate in the following equations, or marginalize over it numerically.

To make progress on Eq.~\eqref{eq:cn_gp_posterior_predictive_integral} requires deriving $\pr(\mu, \sdth^2 \given \ckvecset, \ell)$.
Again, since we know that this posterior has the same functional form as the prior, our task is to determine $\cond\hypm, \cond\hypdisp, \cond\hypdf, \cond\hyptau^2$.
As usual, Bayes' theorem gets the ball rolling:
\begin{align}
  \pr(\mu,\sdth^2 \given \ckvecset, \ell, Q) & \propto \pr(\ckvecset \given \param, Q) \pr(\mu,\sdth^2) \notag \\
  & \propto \pr(\mu,\sdth^2) \prod_n \pr(\inputvec{c}_n \given \param).
\end{align}
Next, by inserting Eqs.~\eqref{eq:cn_iid_with_basis} and~\eqref{eq:ninvchi2_prior},
\begin{align}
  \pr(\mu,\sdth^2 \given & \ckvecset, \ell, Q) \propto \sdth^{-(\hypprior\hypdf+N\nc+3)} \notag \\
  & \times \exp{\frac{-1}{2\sdth^2}\left[\hypprior\hypdf\hypprior\hyptau^2 + \transpose{(\mu-\hypprior\hypm)}\hypprior\hypdisp^{-1}(\mu-\hypprior\hypm)\right]} \notag \\
  & \times \exp{\frac{-1}{2\sdth^2} \sum_n \transpose{(\inputvec{c}_n - \basis\mu)}R_\ell^{-1}(\inputvec{c}_n-\basis\mu)}, \label{eq:mu_sigma_post_messy}
\end{align}
where $R_\ell = r(\kinparvecset, \kinparvecset;\ell)$ and $\basis = \basisfunc^\trans(\kinparvecset)$ are $N\times N$ and $N \times p$ matrices, respectively.
If $\basisfunc(\kinparvec) = 1$, as is assumed for the examples here, then $B = \onevec$, a length $N$ vector of ones.

%and the basis matrix is given by $\basis = \basisfunc^\trans(\kinparvecset) = \onevec$, a length $N$ vector of ones.

So far we can read off the updated value of $\cond\hypdf = \hypprior\hypdf + N\nc$, but must inspect the exponent further for the others.
Define $\ev{\ckvecset}$ as an $N\times 1$ vector that is the average over the $\nc$ orders,
\begin{align}
    \ev{\ckvecset} \equiv \frac{1}{\nc} \sum_n \inputvec{c}_n.
\end{align}
The sum over the quadratic form can be rearranged as
\begin{align}
  \sum_n \transpose{(\inputvec{c}_n - \basis\mu)} & R_\ell^{-1}(\inputvec{c}_n-\basis\mu) \notag \\
  & = \nc \transpose{\left(\ev{\ordervec{\inputvec{c}}_k} - \basis\mu\right)} R_\ell^{-1} \left(\ev{\ordervec{\inputvec{c}}_k} - \basis\mu\right) \notag \\
  & ~~ + \sum_n\transpose{\left(\inputvec{c}_n - \ev{\ordervec{\inputvec{c}}_k}\right)} R_\ell^{-1} \left(\inputvec{c}_n - \ev{\ordervec{\inputvec{c}}_k}\right)
\end{align}
so that we can complete the square for $\mu$ in the exponent of Eq.~\eqref{eq:mu_sigma_post_messy}.
The result is
\begin{align}
  \frac{-1}{2\sdth^2} \left[\cond\hypdf\cond\hyptau^2 + \transpose{(\mu - \cond\hypm)} \cond\hypdisp^{-1} (\mu - \cond\hypm)\right],
\end{align}
where
\begin{align}
  \cond{\hypm} & \equiv \cond{\hypdisp} \left(\hypprior\hypdisp^{-1}\hypprior\hypm + \nc \transpose{\basis}R_\ell^{-1} \ev{\ordervec{\inputvec{c}}_k}\right) \label{eq:hypm_update_corr} \\
  \cond{\hypdisp} & \equiv \left(\hypprior\hypdisp^{-1} + \nc \transpose{\basis} R_\ell^{-1} \basis\right)^{-1} \\
  \cond{\hypdf} & \equiv \hypprior\hypdf + N\nc \\
  \cond\hypdf\cond\hyptau^2 & \equiv \hypprior\hypdf\hypprior\hyptau^2 + \transpose{\hypprior\hypm}\hypprior\hypdisp^{-1}\hypprior\hypm + \sum_n \transpose{\inputvec{c}}_n R_\ell^{-1} \inputvec{c}_n - \transpose{\cond{\hypm}}\cond{\hypdisp}^{-1}\cond{\hypm}. \label{eq:hyptau_update_corr}
\end{align}
%Although $\mu$, $\hypm$, and $\hypdisp$ are scalars in this case, we have left the above equations in vector notation for generality if a more complicated basis were to be used for the mean function as in, e.g., Ref.~\cite{BastosDiagnosticsGaussianProcess2009}.
After some algebra, one can rewrite Eq.~\eqref{eq:hyptau_update_corr} in an intuitive way~\cite{o1994bayesian}:
\begin{align} \label{eq:hyptau_update_corr2}
    \cond\hypdf\cond\hyptau^2 & = \hypprior\hypdf\hypprior\hyptau^2 + \nc s^2 \\
    & + \transpose{(\ev{\ckvecset} - \basis\hypprior\hypm)} \left(\frac{R_\ell}{\nc} + \basis \hypprior\hypdisp
    \transpose\basis\right)^{-1} (\ev{\ckvecset} - \basis\hypprior\hypm), \notag
\end{align}
which updates the scale $\cond\hyptau^2$ by combining prior information with two other terms: one due to the discrepancy between the prior mean and sample mean, and the other due to the sample variance $s^2$ for \emph{correlated} observations,
\begin{align}
    s^2 \equiv \frac{1}{\nc} \sum_n \transpose{(\inputvec{c}_n - \ev{\ckvecset})} R_\ell^{-1} (\inputvec{c}_n - \ev{\ckvecset}).
\end{align}
When $\nc \gg 1$, then $\cond\hyptau^2 \approx s^2$.
If one only wanted point estimates so that the $c_n$ or $\delta\genobs_k$ posteriors remain Gaussian, then $\mu \approx \cond\hypm$ and $\sdth^2 \approx \cond\hyptau^2$ from Eqs.~\eqref{eq:hypm_update_corr} and~\eqref{eq:hyptau_update_corr2} provide approximate Bayesian estimators.

Returning to Eq.~\eqref{eq:cn_gp_posterior_predictive_integral}, we must integrate out $\mu$ and $\sdth^2$ (again, assuming $\ell$ is fixed).
Though the integral over $\mu$ is doable, there is neat trick to see the answer right away: rewrite the GP as a mean $\mu$ (whose distribution is Gaussian) plus a zero-mean GP, that is,
\begin{align}
  c_n(\kinparvec) & = \basisfunc^\trans(\kinparvec)\mu + \epsilon_1(\kinparvec), & \epsilon_1(\kinparvec) & \sim \GP[0, \sdth^2 r(\kinparvec,\kinparvec';\ell)] \label{eq:gp_rewrite_trick} \\
  \mu & = \cond\hypm + \epsilon_2, & \epsilon_2 & \sim \normal(0, \sdth^2\cond\hypdisp). \label{eq:mean_rewrite_trick}
\end{align}
Now, substitute Eq.~\eqref{eq:mean_rewrite_trick} into Eq.~\eqref{eq:gp_rewrite_trick} and use Gaussian sum rules to arrive at
\begin{align} \label{eq:cn_marg_mu}
  c_n(\kinparvec) \given \ckvecset, \sdth^2, \ell \sim \GP{\!\left\{\basisfunc^\trans(\kinparvec)\cond\hypm, \sdth^2{ \left[r(\kinparvec,\kinparvec';\ell) + \basisfunc^\trans(\kinparvec)\cond\hypdisp\basisfunc(\kinparvec')\right]}\right\}},
\end{align}
which has marginalized out $\mu$ for us.
% Above, $\cond\hypdisp$ is added to the correlation element-wise.
The integral over $\sdth^2$, follows from the multivariate analog of Eq.~\eqref{eq:cn_posterior_uncorr}, that is, integrating Eq.~\eqref{eq:cn_marg_mu} over $\sdth^2$ yields a Student-$t$ process $\TP_\nu[m(\kinparvec), \kernel(\kinparvec,\kinparvec')]$~\cite{ShahStudenttProcessesAlternatives2014}
\begin{align} \label{eq:cn_marg_mu_sigma}
  c_n(\kinparvec) \given \ckvecset, \ell \sim \TP_{\cond\hypdf}\!\left\{\basisfunc^\trans(\kinparvec)\cond\hypm, \cond\hyptau^2{ \left[r(\kinparvec,\kinparvec';\ell) + \basisfunc^\trans(\kinparvec)\cond\hypdisp\basisfunc(\kinparvec')\right]}\right\}\!.
\end{align}
Just like GPs, Student-$t$ processes (TPs) are collections of random variables, any finite number of which have a joint $t$ distribution.
Specifically, the density of an $N$ dimensional multivariate $t$ distribution is
\begin{align} \label{eq:mvt_pdf}
  \pr(\mathbf{y}\given\nu, \mathbf{m}, K) & = \frac{|K|^{-1/2}}{(\nu\pi)^{N/2}} \frac{\Gamma(\frac{\nu+N}{2})}{\Gamma(\nu/2)} \notag \\
  & \times{\left(1 + \frac{\transpose{(\mathbf{y} - \mathbf{m})} K^{-1} (\mathbf{y} - \mathbf{m})}{\nu}\right)}^{-\frac{\nu+N}{2}}.
\end{align}
With this notation, $\kappa(\kinparvec,\kinparvec')$ is not the covariance function of $\TP_\nu[m(\kinparvec), \kernel(\kinparvec,\kinparvec')]$, and, similarly, $K$ is not the covariance matrix in Eq.~\eqref{eq:mvt_pdf}.
Rather, the covariance is given by $\nu\kappa(\kinparvec,\kinparvec')/(\nu-2)$ and is only defined for $\nu>2$.
This differs from the notation in, say, Refs.~\cite{BastosDiagnosticsGaussianProcess2009,ShahStudenttProcessesAlternatives2014}.
TPs, like GPs, can be used in regression, etc., but can better handle outliers in the data.
Here, TPs only show up in our convergence model due to our choice of priors, but are convenient objects to work with nevertheless.

For completeness, we provide the conditional distribution for Student-$t$ processes, which is very similar to the GP version in Sec.~\ref{ssub:interpolation_and_regression}.
Suppose $\mathbf{y}_1$ and $\mathbf{y}_2$ are length $n_1$ and $n_2$ vectors, respectively, and
\begin{align}
    \begin{bmatrix}
        \mathbf{y}_1 \\ \mathbf{y}_2
    \end{bmatrix}
    \sim
    t_\nu\left(
    \begin{bmatrix}
        \mathbf{m}_1 \\ \mathbf{m}_2
    \end{bmatrix}
    ,
    \begin{bmatrix}
        K_{11} & K_{12} \\
        K_{21} & K_{22}
    \end{bmatrix}
    \right).
\end{align}
Then
\begin{align} \label{eq:tp_conditional}
    \mathbf{y}_2 \given \mathbf{y}_1 \sim t_{\nu+n_1}{\left(\tilde{\mathbf{m}}_2, \frac{\nu+d_1}{\nu+n_1}\tilde K_{22} \right)},
\end{align}
where $\tilde{\mathbf{m}}_2$ and $\tilde K_{22}$ are defined as in Eqs.~\eqref{eq:interp_mean} and~\eqref{eq:interp_cov}, and $d_1 = \transpose{(\mathbf{y}_1-\mathbf{m}_1)} K_{11}^{-1} (\mathbf{y}_1-\mathbf{m}_1)$.
One could derive this by starting with $\mathbf{y} \given \sigma^2 \sim \normal(\mathbf{m}, \sigma^2 K)$ and $\sigma^2 \sim \invchisq(\nu, 1)$, using the Gaussian conditional rules described in Sec.~\ref{ssub:interpolation_and_regression}, and then marginalizing over $\sigma^2$ at the end.

It is important to note that Eq.~\eqref{eq:tp_conditional} is \emph{not} how we compute conditionals for, e.g., Eq.~\eqref{eq:cn_marg_mu_sigma}.
This is because conditioning and marginalizing over the mean do not commute.
Consider the case where $\mathbf{f} = \transpose{\begin{bmatrix} \mathbf{f}_1 & \mathbf{f}_2 \end{bmatrix}}$ is normally distributed just as $c_n$.
Then
\begin{align} \label{eq:gp_marg_conditional_integral}
    \pr(\mathbf{f}_2 \given \mathbf{f}_1) & = \int \pr(\mathbf{f}_2 \given \mathbf{f}_1, \mu, \sdth^2) \pr(\mu, \sdth^2 \given \mathbf{f}_1) \dd{\mu} \dd{\sdth^2}
\end{align}
from which it is clear that one must first follow the conditional rules of Eqs.~\eqref{eq:interp_mean}--\eqref{eq:interp_cov} and also update the posterior of $\mu$ and $\sdth^2$ before integrating them out.
Thus, following the same logic as above,
\begin{align}
    \pr(\mathbf{f}_2 \given \mathbf{f}_1) = t_{\cond\nu}{\left[R_{21}R_{11}^{-1}\mathbf{f}_1 + \tilde \basis_2 \cond\eta, \cond\tau^2 (\tilde R_{22} + \tilde \basis_2 \cond\hypdisp \transpose{\tilde \basis_2})\right]}, \label{eq:gp_marg_conditional}
\end{align}
where, remembering that $\basis = \onevec$ (a vector of ones) is the basis matrix for our coefficients,
\begin{align} \label{eq:conditional_basis_matrix}
    \tilde \basis_2 \equiv \basis_2 - R_{21}R_{11}^{-1} \basis_1.
\end{align}
Of course, there could be other \iid\ data beyond just $\mathbf{f}_1$ that could be used to compute $\pr(\mu, \sdth^2 \given \mathbf{f}_1)$, which would serve only to adjust $\cond\hypm$, $\cond\hypdisp$, $\cond\hypdf$, and $\cond\hyptau^2$.

The above analysis is sufficient to finish our derivation for full uncertainty quantification of $\genobs_k$, though the details will vary depending on the application as described in Sec.~\ref{sub:application_types}.
In all cases, the derivation involves simply using $\muth$ and $\covth$ from Sec.~\ref{sub:application_types} as the mean and covariance for the Gaussian $\pr(\mathbf{f}_2 \given \mathbf{f}_1, \mu, \sdth^2)$ in Eq.~\eqref{eq:gp_marg_conditional_integral}, and then integrating out $\mu$ and $\sdth^2$.
Below we provide the mean and scale function for $\TP_\hypdf[\muth(\kinparvec), \covth(\kinparvec,\kinparvec')]$ in each case.

\textbf{Inexpensive Predictions.}
Let primes denote means with $\mu \to \hypm$, e.g.,
\begin{align}
    \discrmean{k}'(\kinparvec) & \equiv \discrbasisfunc{k}^\trans(\kinparvec)\hypm = \genobsref(\kinparvec)\frac{Q(\kinparvec)^{k+1}}{1 - Q(\kinparvec)}\hypm.
\end{align}
Then
\begin{align}
    \muth(\kinparvec) & = \genobs_k(\kinparvec) + \discrmean{k}'(\kinparvec) \label{eq:tp_muth_inexpensive} \\
    \covth(\kinparvec, \kinparvec'; \ell) & = \cond\hyptau^2 \left[\discrcorr{k}(\kinparvec, \kinparvec'; \ell) + \discrbasisfunc{k}^\trans(\kinparvec)\cond\hypdisp\discrbasisfunc{k}(\kinparvec') \right]. \label{eq:tp_covth_inexpensive}
\end{align}

\textbf{Expensive Predictions.}
Define $\tilde\basisfunc_i(\kinparvec)$ for $i \in \{k, \delta k\}$ as the non-vectorized version of Eq.~\eqref{eq:conditional_basis_matrix}, using the appropriate $\basisfunc_i(\kinparvec)$ and $\genobscorr{i}$.
Then
\begin{align}
    \muth(\kinparvec) & = \tildegenobsmean{k}'(\kinparvec) + \discrmean{k}'(\kinparvec) \label{eq:tp_muth_expensive} \\
    \covth(\kinparvec, \kinparvec'; \ell) & = \cond\hyptau^2 \Big\{\tildegenobscorr{k}(\kinparvec, \kinparvec'; \ell) +  \discrcorr{k}(\kinparvec, \kinparvec'; \ell) \notag \\
    & \hspace{0in} + \big[\tildegenobsbasisfunc{k}(\kinparvec)+\discrbasisfunc{k}(\kinparvec) \big]^\trans \cond\hypdisp \big[\tildegenobsbasisfunc{k}(\kinparvec')+\discrbasisfunc{k}(\kinparvec') \big]\Big\}. \label{eq:tp_covth_expensive}
\end{align}

\textbf{Constraints.}
One only needs to make the replacements $\discrmean{k}' \to \tildediscrmean{k}'$, $\discrbasisfunc{k} \to \tildediscrbasisfunc{k}$ and $\discrcorr{k} \to \tildediscrcorr{k}$ in the above equations.

In the final part of this appendix, we derive the formulas for the posterior $\pr(\ell, \inputvec{Q} \given \genobsvecset) \propto \pr(\genobsvecset \given \ell, \inputvec{Q}) \pr(\ell, \inputvec{Q})$.
The quantity $\pr(\genobsvecset \given \ell, \inputvec{Q})$ is (almost) the marginal likelihood (or, evidence), and can be computed exactly.
The first step is to make a change of variables from the data $\genobsvecset$ to coefficients
\begin{align} \label{eq:genobsvec_given_ell_Q}
  \pr(\genobsvecset\given\ell, \inputvec{Q}) = \frac{\pr(\ordervec{\inputvec{c}}_k\given\ell)}{\prod_{n,i} \abs{\genobsref(\kinparvec_i)Q^n(\kinparvec_i)}}.
\end{align}
To find $\pr(\ordervec{\inputvec{c}}_k\given\ell)$, we will make use of the same normalizing constant trick as in the uncorrelated case.
Again, let primes denote unnormalized distributions and $Z_{\text{dist.}}$ be their normalizing constants. 
By Bayes' theorem,
\begin{align}
  \pr(\mu, \sdth^2 \given \ordervec{\inputvec{c}}_k, \ell) & = \frac{Z_{\chi,0}^{-1}Z_{\normal}^{-\nc}}{\pr(\ordervec{\inputvec{c}}_k\given\ell)} \ninvchisq'(\hypprior\hypm,\hypprior\hypdisp,\hypprior\hypdf,\hypprior\hyptau^2) \notag \\ & ~~ \prod_n \normal'(\mu,\sdth^2R_\ell)
\end{align}
and by conjugacy
\begin{align}
  \pr(\mu, \sdth^2 \given \ordervec{\inputvec{c}}_k, \ell) = \frac{1}{Z_{\chi}} \ninvchisq'(\cond\hypm, \cond\hypdisp, \cond\hypdf, \cond\hyptau^2).
\end{align}
Given that
\begin{align}
  Z_{\chi,0}^{-1} & = \frac{(\hypprior\hypdf\hypprior\hyptau/2)^{\hypprior\hypdf/2}}{\sqrt{|2\pi\hypprior\hypdisp|} \Gamma(\hypprior\hypdf/2)},
  &
  Z_{\normal}^{-1} & = \frac{1}{\sqrt{|2\pi R_\ell|}}
\end{align}
with a similar form for $Z_{\chi}$, then
\begin{align} \label{eq:ckvec_given_ell}
  \pr(\ordervec{\inputvec{c}}_k\given\ell) %= \frac{\cond{Z_{\ninvchisq}}}{Z_{\ninvchisq} Z_{\normal}^{\nc}}
  = \frac{\Gamma(\cond\hypdf/2)}{\Gamma(\hypprior\hypdf/2)} \sqrt{\frac{|\cond{\hypdisp}|/|\hypprior\hypdisp|}{|2\pi R_\ell|^{\nc}} \frac{(\hypprior\hypdf\hypprior\hyptau^2/2)^{\hypprior\hypdf}}{(\cond\hypdf\cond\hyptau^2/2)^{\cond\hypdf}} }.
\end{align}
The joint posterior $\pr(\ell, \inputvec{Q} \given \genobsvecset)$ follows from Eqs.~\eqref{eq:genobsvec_given_ell_Q}, \eqref{eq:ckvec_given_ell} and a choice of prior $\pr(\ell, \inputvec{Q})$.
It follows that the unnormalized marginal posteriors for $\inputvec{Q}$ and $\ell$ are
\begin{align}
  \pr(\inputvec{Q} \given\genobsvecset, \ell) & \propto \frac{\pr(\inputvec{Q})}{\cond\hyptau^{\cond\hypdf}\prod_{n,i} |Q^n(\kinparvec_i)|} \label{eq:Q_given_ell_posterior_corr} \\
  \pr(\ell\given\genobsvecset, \inputvec{Q}) & \propto \frac{\pr(\ell)}{\cond\hyptau^{\cond\hypdf}} \sqrt{\frac{|\cond{\hypdisp}|}{|R_\ell|^{\nc}}}. \label{eq:ell_given_Q_posterior_corr}
\end{align}
The posterior for the breakdown scale $\Lambda_b$, where $Q(\kinparvec;\Lambda_b) = f(\kinparvec)/\Lambda_b$, follows from the same logic as in Appendix~\ref{sub:pointwise_model} and is given by
\begin{align}
    \pr(\Lambda_b \given \genobsvecset, \ell, \inputvec{f}) & \propto \frac{\pr(\Lambda_b)}{\cond\hyptau^{\cond\hypdf}\prod_{n,i} |Q^n(\kinparvec_i)|}. \label{eq:breakdown_scale_posterior_corr}
\end{align}
MAP values can then be found numerically and used as point estimates in the analytic equations derived in this Appendix.

% \section{Algorithm}

% \IncMargin{1em}
% \begin{algorithm}
% \DontPrintSemicolon
% % \SetKwData{Left}{left}\SetKwData{This}{this}\SetKwData{Up}{up}
% % \SetKwFunction{Union}{Union}\SetKwFunction{FindCompress}{FindCompress}
% % \SetKwInOut{Input}{input}\SetKwInOut{Output}{output}
% \SetInd{0.5em}{1em}
% \SetNlSkip{0.5em}
% % \SetAlgoHangIndent{20em}
% \KwData{Order-by-order predictions $\genobsvecset$.}
% \KwResult{Full posterior $\pr(\genobs \given \genobsvecset, I)$}
% \BlankLine
% Choose theoretically motivated values of $\genobsref$ and $Q$\;
% Extract coefficients $c_n$ using Eq.~?\;
% Assess which coefficients are relevant for predicting future coefficients, toss the others\;
% Choose a mean and covariance function, along with $\hypprior\hypm$, $\hypprior\hypdisp$, $\hypprior\hypdf$, and $\hypprior\hyptau^2$\;
% Split the data into training and validation data\;
% Compute $\pr(\param \given \ckvecset)$\;
% Extract the mean and covariance of the emulator\;
% Compute the diagnostics of the emulator against the validation data\;
% Assess results\;
% \caption{Truncation Error}\label{algo_truncation}
% \end{algorithm}\DecMargin{1em}

%%%%%%%%%%%%%%%%%%%%%%%%%%%%%%%%%%%%%%%%%%%%%%%%%%%%%%%%%%%%%%%

\clearpage
\bibliography{EMN_Correlations_Refs}

\end{document}